\documentclass[aps,prb,twocolumn,superscriptaddress,floatfix,10pt,letterpaper,showpacs]{revtex4-1}

\pdfoutput=1

\begin{document}

\begin{titlepage}

\title{Symmetry Enforced Non-Abelian Topological Order at the Surface of a Topological Insulator}
\author{Xie Chen}
\affiliation{Department of Physics, University of California, Berkeley, CA 94720, USA.}
\author{Lukasz Fidkowski}
\affiliation{Department of Physics and Astronomy, Stony Brook University, Stony Brook, NY 11794-3800, USA.}
\author{Ashvin Vishwanath}
\affiliation{Department of Physics, University of California, Berkeley, CA 94720, USA.}

\affiliation{Materials Science Division, Lawrence Berkeley National Laboratories, Berkeley, CA 94720, USA.}

\begin{abstract}
The surfaces of three dimensional topological insulators (3D TIs) are generally described as Dirac metals, with a single Dirac cone. It was previously believed that a gapped surface implied breaking of either time reversal $\mathcal T$  or $U(1)$ charge conservation symmetry. Here we discuss a novel possibility in the presence of interactions, a surface phase that preserves all symmetries but is nevertheless gapped and insulating. Then the surface must develop topological order of a kind that cannot be realized in a 2D system with the same symmetries.  We discuss candidate surface states  - non-Abelian Quantum Hall states which, when realized in 2D, have $\sigma_{xy}=1/2$ and hence break $\mathcal T$ symmetry. However, by constructing an exactly soluble 3D lattice model, we show they can be realized as $\mathcal T$ symmetric surface states. The corresponding 3D phases are confined,  and have  $\theta=\pi$ magnetoelectric response.
Two candidate states have the same 12 particle topological order, the (Read-Moore) Pfaffian state with the neutral sector reversed, which we term T-Pfaffian topological order, but differ in their $\mathcal T$ transformation.  Although we are unable to connect either of these states directly to the superconducting TI surface, we argue that one of them describes the 3D TI surface, while the other differs from it by a bosonic topological phase. We also discuss the 24 particle Pfaffian-antisemion topological order (which can be connected to the superconducting TI surface) and demonstrate that it can be realized as a $\mathcal T$ symmetric surface state.
\end{abstract}


\maketitle

\end{titlepage}



\section{Introduction}
\label{intro}

Three dimensional topological insulators host unusual surface states that can be described in a number of different ways\cite{Hasan2010,Qi2011,Hasan2011}. In models of free electrons that respect time reversal and charge conservation symmetries which are necessary to describe this phase, the surface is metallic.  The surface electronic structure is comprised of an odd number of 2D Dirac cones, which is impossible to realize in a purely 2D system with time reversal invariance. It is crucial  that electrons transform as Kramers pairs i.e.  time reversal $\mathcal T$ acting twice on electrons  gives ${\mathcal T}^2=-1$. 

Other surface terminations of the topological bulk are also interesting. If $\mathcal T$ is broken only at the surface, the metallic edge can be gapped to yield an insulating surface. The topological bulk properties are revealed in the properties of  domain walls between opposite $\mathcal T$ breaking regions on the surface. The domain walls are necessarily metallic and host an odd number of chiral Dirac fermions $c_-=2n+1$\cite{Fu2007}.  Thus, each domain is associated with  Hall conductance $\sigma_{xy}=n+\frac12$, (and $\kappa_{xy}/T = n+\frac12$ in the natural units for thermal Hall conductance) where $n$ is an integer.  Such a half-integer Hall conductance if forbidden in a 2D system in the absence of electron fractionalization and is directly related to the magnetoelectric effect\cite{Qi2008,Essin2009}. The magnetoelectric polarizability of Topological insulators is $\theta=\pi$ \cite{} in contrast to trivial, time reversal symmetric insulators which have $\theta=0$. If, on the other hand we break charge conservation on the surface, by pairing and condensing Cooper pairs, the resulting surface superconductor also has exotic properties - its vortices host Majorana zero modes\cite{Fu2008}, which only occur in 2D systems if $\mathcal T$ symmetry is broken. 

It was believed that these were the only possible surface states of the 3D topological insulator  - i.e. they must either be gapless (e.g. metallic) or break one of the defining symmetries. Recently, inspired by the study of bosonic topological phases\cite{Chen2011,Chen2012a, Levin2012,Lu2012a,Vishwanath2013,Metlitski2013}, and fermionic topological superconductors \cite{Fidkowski2013}, a different surface termination has been recognized, that preserves all symmetries and develops an energy gap at the surface\cite{Vishwanath2013,Wang2013,Burnell2013,Metlitski2013,Fidkowski2013}. In this situation the surface develops topological order i.e. there are anyonic excitations bound to the surface. While the topological order itself may be realized in 2D, the transformation properties of anyons under action of the symmetry is unlike in any 2D system. 

A close analog of this problem was recently discussed in the context of fermionic topological superconductors protected by $\mathcal T$ symmetry, where a non-Abelian surface topological order was identified\cite{Fidkowski2013}.  The discussion in this paper will closely follow this earlier work. Of course, such surface states will only be realized in systems with strong electron correlations - and represents a qualitatively new property of interacting topological insulators. This is analogous to the argument that in a 2D  quantum spin system with S=1/2 per unit cell, a gapped, symmetric state must be topologically ordered\cite{Hastings2004}. Indeed the absence of ordering is taken as a sign of topological order. Similarly, if on the surface of a 3D topological insulator, no sign of superconductivity or $\mathcal T$ breaking is present but an energy gap opens, this will be indicative of topological order.  In fact, as discussed in the context of fermionic topological superconductors\cite{Fidkowski2013}, and as we will see below, the topological order here is required to be {\em  non-Abelian}.  This mechanism may provide a route to realizing non-Abelian topological order, which would provide further impetus in the search for strongly correlated topological insulators such as SmB$_6$ \cite{Wolgast,Xia,Zhang2013}.

Let us discuss some of the key physical requirements that the topologically ordered surface of a 3D TI should satisfy. (a) `$Z_2$ ness': Given the $Z_2$ classification of free fermion TIs, a pair of topologically ordered surfaces should `unwind', and should be connected to a trivial confined phase. (b) Magnetoelectric response $\theta=\pi$: In the absence of topological order, surface domains of the $\mathcal T$ breaking insulating surface  can be assigned a Hall conductivity $\sigma_{xy}= 1/2$  and thermal Hall conductivity $\kappa_{xy}/T=1/2$ (modulo integers).  These are equivalent to the statement that the magnetoelectric response is $\theta=\pi$. The implication for  the topologically ordered states, is that in their 2D versions, which break $\mathcal T$, should have $\sigma_{xy}=1/2$ and $\kappa_{xy}/T=1/2$ modulo integers. (c) On breaking charge conservation symmetry, it should be possible to remove the topological order while preserving $\mathcal T$. The resulting superconductor should host hc/2e vortices with Majorana zero modes in their core. 

At first, a promising choice appears to be the Read-Moore Pfaffian state\cite{Moore1989}. One picture of this phase is to consider a topological superconductor of spin polarized electrons in a $p_x+ip_y$ pairing state, while the Cooper pairs form a $\nu=1/8$ bosonic Laughlin state. The latter has anyon excitations with fractional charge $q_k=2e\frac{k}8$ and a charged chiral edge state. The bound state of the superconductor quasiparticle and the charge $e$ excitation is identified as the electron and only those excitations that braid trivially with the electron are retained. This phase has 12 quasiparticles (including the electron) and $\sigma_{xy}=\frac{(2e)^2}{8h}=\frac12\frac {e^2}h$. While this satisfies one of our criteria, it is readily seen from the topological spins of the associated topological order that this cannot be realized in a $\mathcal T$ invariant way even on a surface. 

However, a simple modification produces a more promising candidate which we call the T-Pfaffian ($\mathcal T$ preserving Pfaffian) \footnote{This state was independently identified by P. Bonderson, X. Qi, C. Nayak (arXiv:1306.3230 (2013))}. Consider the time reversed superconductor ( $p_x-ip_y$ ) combined in the identical way with the Abelian topological order of Cooper pairs. This theory also has $\sigma_{xy}=1/2$ and $\kappa_{xy}/T=1/2$ when realized in 2D as required.  Moreover, the topological spins of the quasiparticles (Table \ref{T_spin}) now appear compatible with time reversal symmetry. We construct an exactly soluble 3D model that explicitly demonstrates that this state can be realized on the surface of a 3D bulk system while retaining $\mathcal T$ and charge U(1) symmetries. Since the surface topological order is forbidden in a strictly 2D system, we have realized {\em  a} 3D Topological Insulator. Moreover, this phase has magnetoelectric response $\theta=\pi$. The remaining question is - is this the same phase as the free fermion topological insulator? 

A necessary condition to make this identification is that in the absence of charge conservation symmetry (i.e. induced by proximity coupling the surface to a superconductor), one should recover the  TI surface superconductor, without topological order, but with Majorana zero modes in the vortex cores. The T-Pfaffian state however allows no simple way to exit the topological phase even when charge conservation is absent, while retaining $\mathcal T$. On the other hand, we find that there are two versions of the T-Pfaffian state (T-Pfaffian$_\eta$, with $\eta=\pm1$) which differ in the way the nonabelian particles transform under time reversal symmetry. The  non-abelian particles with bosonic (fermionic) topological spin is assigned $\mathcal T^2=\eta$  ($\mathcal T^2=-\eta$).  We can then demonstrate the following fact: (i) The two states corresponding to $\eta=+1$ and $\eta=-1$ differ by the surface topological order of a bosonic topological superconductor (BTSc) with a ${\mathcal Z}_2$ classification and (ii) they can differ from the 3D free fermion TI surface at most, by the same BTSc surface topological order.  This  implies that one of them {\em must be} the free fermion TI surface state while the other represents a mixture with a BTSc although, unfortunately, we cannot  at present specify which of the $\eta=\pm1$ is the 3D TI surface.


We also discuss a second topological order, the Pfaffian-antisemion state obtained recently \cite{Wang2013a,Metlitski2013a} by a series of elegant physical arguments. This state is a tensor product of the Read-Moore Pfaffian state as discussed above, with a neutral anti-semion theory $\{1,\bar{s}\}$ and has 24 quasiparticles. Its statistics is compatible with $\mathcal T$ symmetry, and furthermore passes the necessary requirements for being identified with the 3D TI surface state, including realizing the TI surface superconductor on breaking charge conservation \cite{Wang2013a, Metlitski2013a}. Here we prove that it is indeed realizable as the surface state of a 3D bulk system with the requisite symmetries, by constructing an exactly soluble 3D lattice model. In both this and the T-Pfaffian case, we find time reversal symmetry is respected {\em only} if the electrons transform projectively, i.e. as ${\mathcal T}^2=-1$.

A central tool will be the Walker-Wang construction\cite{Walker2012,Keyserlingk2013,Burnell2013,Fidkowski2013} of an exactly soluble lattice Hamiltonian, that realizes the desired surface topological order, while maintaining a topologically trivial bulk\footnote{This is strictly true when the surface topological order is modular, and contains anyons with nontrivial mutual statistics with at least one other particle}. The model realizes a ground state wave function which is a superposition of 3D loops, one for each quasiparticle of the surface topological phase. The amplitude for any configuration is obtained as follows. The topological order is represented in terms of the $R$ and $F$ symbols that are associated with certain basic loop moves\cite{Kitaev2006}. First, one  projects the loop configuration onto a 2D plane, and relates it to a reference configuration using the elementary moves. For each move, the amplitude acquires a factor that is related to the $R$ and $F$ symbols. Based on the resulting state one can readily show that  an anyon that has nontrivial mutual statistics with some other excitation, is necessarily confined to the surface. Furthermore, the surface excitations realize the required topological order. A minor caveat here is that for simplicity we work with bosonic Walker Wang models, without elementary fermions in the Hilbert space\cite{Fidkowski2013}. Since the surface topological order is non-modular, i.e. it contains a fermion excitation that has trivial mutual statistics with everything else, this excitation is deconfined in the bulk. Hence the 3D state that is realized is a Z$_2$ gauged version of the topological insulators, i.e. it involves bulk fermions that carry gauge charge and loops carrying $\pi$ gauge flux. This state may thought of as being obtained from a bosonic model, from a parton construction $b=f_\uparrow f_\downarrow$, where the Z$_2$ gauge charged fermions $f_{\uparrow,\downarrow}$  are placed in a 3D topological phase\cite{Swingle2011}. This can be readily rectified by introducing elementary fermions $c_\sigma$ and condensing the product $c^\dagger_\sigma f_\sigma$ which confines the gauge flux and removes the bulk topological order.

Before moving to the technical details we raise the following question that may have occurred to some readers. How is the $\theta=\pi$ magneto electric response of 3D topological insulator reconciled with the topologically ordered insulating surfaces with $\mathcal T$ symmetry? Note, $\theta=\pi$ implies that a weak applied magnetic field produces a surface charge density of $e/2$ (modulo integer multiples of $e$) per flux quantum. Of course, such a charge density is meaningful only if the surface is also insulating. Also, charge is only determined modulo $e$ since integer multiples of $e$ can be screened by surface electrons. The fractional part however cannot be screened by electrons and is a bulk property. Specifying whether this charge density is $\pm e/2$ leads to the usual argument for breaking of time reversal symmetry at the surface.   The key observation here is that the candidate states both contain charge $e/2$ excitations which can screen the induced charge, and hence breaking of $\mathcal T$  symmetry is not required. 

\section{The T-Pfaffian Topological Order}
First let us introduce the topological order in the T-Pfaffian state, including the anyons types and their fusion and braiding rules. 

The T-Pfaffian state is a twisted version of the Pfaffian state (such that the state could potentially be time reversal invariant). Similar to Pfaffian, it is the combination of the non-Abelian Ising theory with the Abelian $U(1)_8$ theory. The Ising theory describes a gauged a $p+ip$ superconductor with $Z_2$ fluxes and contains anyons labeled by $I$,$\si$,$\psi$. $I$ labels the trivial vacuum, $\psi$ is the fermion in the superconductor and $\si$ is the $Z_2$ flux. They fuse according to
\begin{eqnarray*}
\si \times \si &=& I + \psi\\ \si \times \psi&=&\si \\ \psi \times \psi&=&I
\end{eqnarray*}
$\si$ is hence nonabelian and has quantum dimension $\sqrt{2}$ while $\psi$ has quantum dimension $1$. The topological spins for the anyons are
\be
\theta_I=1, \theta_{\si}=e^{-i\frac{\pi}{8}}, \theta_{\psi}=-1
\ee
The braiding of the fermion $\psi$ around the $Z_2$ flux results in a phase factor of $-1$. The Ising sector is neutral and does not carry charge.

The $U(1)_8$ theory describes the $\nu=\frac{1}{2}$ quantum Hall state where charge $2e$ electron pairs form an effective $\nu=\frac{1}{8}$ bosonic Laughlin state. The Chern-Simons effective theory for this state is
\be
\mathcal{L}=\frac{8}{4\pi}a_{\mu}\partial_{\nu}a_{\lambda}\epsilon^{\mu\nu\lambda}
\ee
There are $8$ abelian anyons in the theory labeled by $k=0,1,2,...,7$ which add in the usual way when they fuse
\be
k_1\times k_2= (k_1+k_2)\ \text{mod}\ 8
\ee
The topological spins for the anyons are given by
\be
\theta_k=e^{i\frac{\pi}{8}k^2}
\ee
A full braiding of particle $k$ around particle $k'$ results in a phase factor of
\be
\theta_{kk'}=e^{i\frac{\pi}{4}kk'}
\ee
The $U(1)_8$ sector carries fractional charge with the $k$ particle carrying $\frac{ke}{4} \text{mod} 2e$ charge. 

The T-Pfaffian theory is then obtained by combining $I$ and $\psi$ of the Ising theory with even $k$ of the $U(1)_8$ theory and $\si$ with odd $k$. That is, the T-Pfaffian theory has anyons
\be
\begin{array}{llll}
I_0 & I_2 & I_4 & I_6 \\ \nonumber
\psi_0 & \psi_2 & \psi_4 & \psi_6 \\ \nonumber
\si_1 & \si_3 & \si_5 & \si_7
\end{array}
\ee
The charge assignment of the T-Pfaffian theory is carried over from the $U(1)_8$ theory. The fusion and braiding statistics of the combined anyons is the product of the fusion and braiding of the Ising part and the $U(1)$ part. In particular, the topological spin for the combined anyons are
\begin{table}[t]

\be
\begin{array}{|c|c|c|c|c|c|c|c|c|}
\hline
     & 0 & 1 & 2 & 3 & 4 & 5 & 6 & 7 \\ \hline
I    & 1 &   & i &   & 1 &   & i &   \\ \hline
\si  &   & \ 1 &   & -1&   & -1&   & \ 1 \\ \hline
\psi & -1&   & -i&   & -1&   & -i&  \\ \hline 
& & & & & & & &\\ \hline 
\rm Charge &0 & e/4 &  e/2 & 3e/4 & e & 5e/4 & 3e/2 & 7e/4 \\\hline
\end{array}
\ee

\label{T_spin}

\caption{Topological spin and electrical charge assignments of T-Pfaffian state. Excitations are labeled by $X_k$ where $X\in \{1,\,\sigma,\, \psi\}$ are the rows  and $k\in \{0,\,1,\dots,7\}$ are columns. Charge assignments are at the bottom of the table. The particle $\psi_4$ is the electron - a charge e fermion  with trivial mutual statistics with all other particles. }

\end{table}
The quantum dimensions of the $I_k$ and $\psi_k$ (even $k$) anyons are $1$ and that of the $\si_k$ (odd $k$) anyons are $\sqrt{2}$. 
The particle and anti-particle pairs in the T-Pfaffian theory are
\be 
I_k \sim I_{(8-k) \text{mod}\ 8}, \psi_k \sim \psi_{(8-k)\text{mod}\ 8}, \si_k \sim \si_{(8-k)\text{mod}\ 8}
\ee
Obviously $I_0$ is the vacuum in the theory. Moreover, it is easy to check that a full braiding of $\psi_4$ around any other particle in the theory leads to a phase factor of $1$. That is, the $\psi_4$ particle is a local excitation of the system and cannot be seen by a braiding operation far away. In fact, $\psi_4$ has topological spin of $-1$, carries charge $e$ and is therefore the electron in the system.

Note that while the T-Pfaffian state has the same anyon types as the Pfaffian state, the statistics of the two are different. In particular, the the statistics of the Ising part is taken to be the complex conjugate of that in Pfaffian. Therefore, for example, the topological spin of the $\si_1$ anyon is $1$ in the T-Pfaffian theory while it is $e^{i\frac{\pi}{4}}$ in the Pfaffian theory.

\section{Time reversal symmetry on T-Pfaffian}

Is it possible to realize the T-Pfaffian topological order in a time reversal invariant system? 

For pure 2D system, the answer is no. This is easy to see if we look at the edge of the system. The Ising part of the theory has chiral central charge $c_{-}=-\frac{1}{2}$ and is neutral, hence does not contribute to charge Hall conductance $\si_{xy}$. On the other hand, the $U(1)_8$ part of the theory has chiral central charge $c_{-}=1$ and $\si_{xy}=\frac{1}{2}$. Therefore, the T-Pfaffian theory has total chiral central charge $c_-=\frac{1}{2}$ and charge Hall conductance $\si_{xy}=\frac{1}{2}$. Obviously this is not possible in a pure 2D system with time reversal symmetry. 

However, such argument fails if the T-Pfaffian theory is realized on the surface of a 3D gapped system, because the 2D surface of a 3D system does not have an edge of its own. Therefore, the chiral edge of the T-Pfaffian state no longer presents an obstruction to realization in a time reversal invariant system. On the surface of a 3D system, we can only probe how time reversal symmetry acts on the excitations in the bulk of the 2D system, namely the anyons. First, the action of time reversal symmetry takes complex conjugation of all the braiding and fusion processes among the anyons. Moreover, time reversal symmetry can map one anyon type to another. Therefore, if the T-Pfaffian theory can be realized on the surface of a 3D time reversal invariant system, then the time reversal symmetry must exchange the anyon types in such a way that the fusion and braiding amplitudes are invariant under both complex conjugation and this exchange of anyon types.

Such an exchange of anyon type does seem to exist if we consider the topological spins of the anyons, which describe the self rotating processes of the anyons. From Eqn.\ref{T_spin}, we can see that the topological spins remain invariant if the time reversal symmetry performs the following exchange
\be
I_2 \leftrightarrow \psi_2, \ I_6 \leftrightarrow \psi_6
\ee
together with taking complex conjugation. Compare to the Pfaffian state where such an exchange of anyon type, hence time reversal symmetry, cannot exist. In particular, the topological spins for the four $\si$ anyons in the Pfaffian theory are
\be
\theta_{\si_1}=e^{i\frac{\pi}{4}}, \theta_{\si_3}=e^{i\frac{5\pi}{4}}, \theta_{\si_5}=e^{i\frac{5\pi}{4}}, \theta_{\si_7}=e^{i\frac{\pi}{4}},
\ee
which do not form time reversal invariant pairs.

Moreover, we can check that the exchanges $I_2$ with $\psi_2$ and $I_6$ with $\psi_6$ is consistent with the fusion rules of the T-Pfaffian theory. For example, the fusion process of
\be
I_2 \times \psi_4 = \psi_6
\ee
is mapped to
\be
\psi_2 \times \psi_4 = I_6
\ee
under this exchange, which is again a valid fusion process in the T-Pfaffian theory.

Therefore, we have found an exchange of anyon types which, together with complex conjugation, keeps the fusion rules and the topological spins of the T-Pfaffian theory invariant. 

\section{Local time reversal symmetry action}
\label{local_T}

The exchange of anyon types, however, does not completely specify the action of time reversal symmetry on the T-Pfaffian state. In particular, for anyon types which do not change under time reversal, one can ask whether time reversal acts as $T^2=1$ or $-1$ on the anyon locally. This is a legitimate question to ask, because away from the anyonic excitations, the state remains invariant under time reversal. If anyon types does not change under time reversal, then time reversal acts effectively locally around the anyon. As shown in \onlinecite{Levin2012a}, the local action of time reversal can only square to $1$ or $-1$. If $T^2=-1$ on a particular anyon, then then anyon has an extra spin label and leads to a local $2$ fold degeneracy under time reversal symmetry when it is separated from all other anyons. On the other hand, if the anyon changes type under time reversal, then the effective action of time reversal is nonlocal on the state and it is not well defined to talk about $T^2$ locally for the anyon.

The local time reversal symmetry action represented by this $T^2=\pm 1$ information is important because with different local action, the topological state can correspond to totally different bulk phase. For example, the statistics of $Z_2$ gauge theory with anyons $\{I,e,m,\epsilon\}$ is time reversal invariant with no exchange of anyons. If the two bosonic particles $e$ and $m$ both transform as $T^2=1$, the state can be realized in 2D time reversal invariant system. However, if they both transform as $T^2=-1$, the state can only be realized on the surface of 3D bosonic topological superconductors\cite{Vishwanath2013,Wang2013}. For the T-Pfaffian state, we would be interested in the $T^2$ transformation for the charged boson $I_4$, the chargeless fermion $\psi_0$, the electron $\psi_4$ and all the nonabelian $\si_k$ particles. In particular, the $T^2$ transformation of $\psi_4$ would tell us whether we are dealing with a $T^2=1$ or $-1$ topological insulator.

In this section, we state the general rules for determine the $T^2$ information for anyons. We give the motivation for setting these rules and apply them to the T-Pfaffian state. In appendix \ref{FR_T}, we will provide an algebraic proof for these rules in terms of the exactly solvable Walker-Wang model realizing the particular topological state on the surface.

The local action of time reversal on the anyons can be determined from the following two rules that applies to both Abelian and non-Abelian anyons:
\begin{itemize}
\item{{\bf Rule 1:} If anyons $i$ and $j$ fuse into $k$ and none of $i$, $j$, $k$ change type under time reversal, then $T^2_k=T^2_i \times T^2_j$.}
\item{{\bf Rule 2:} If $i$ maps into $\bar{i}$ (different from $i$) under time reversal and the braiding of $i$ around $\bar{i}$ in the fusion channel $k$ resulting in a phase factor of $1(\text{or\ }-1)$, then $T^2_k=1(\text{or\ }-1)$.}
\end{itemize} 

The first rule comes from considering a region with two anyons $i$ and $j$. If $i$ and $j$ are separated far enough (larger than correlation length), then they each have well defined $T^2$. The total local time reversal symmetry action on the whole region is composed of that on $i$ and $j$ separately. However, for an observer very far away from this region, the total anyonic charge in the region is $k$. Therefore, $T^2_k=T^2_i\times T^2_j$. Note that one consequence of this rule is that a particle and its antiparticle have the same $T^2$.

The second rule comes from considering a region with $i$ and $\bar{i}$. Applying time reversal, $i$ changes into $\bar{i}$ and vice versa. This is equivalent to rotating the $(i,\bar{i})$ pair by $180$ degree. Applying time reversal twice, we have rotated the pair by $360$ degree which is equivalent to a full braiding of $i$ around $\bar{i}$ and results in a phase factor of $1(\text{or\ }-1)$. However, for an observer far away from this region, the total anyonic charge in the region is $k$ and the $1(\text{or\ }-1)$ phase factor comes from $T^2_k$.

Applying these two rules, we can find out possible $T^2$ action on each anyon locally. There may be more than one set of $T^2$ values for all the anyons satisfying these rules.

Let's now apply these rules to T-Pfaffian and determine the local time reversal symmetry action. First of all, the electron $\psi_4$ is the fusion product of $I_2$ and $\psi_2$ which map into each other under time reversal. Because the braiding of $I_2$ around $\psi_2$ gives a $-1$, $\psi_4$ transform as $T^2=-1$. Therefore, we are indeed dealing with a $T^2=-1$ topological insulator. Next, because $\si_1$ and $\si_3$ can fuse both to $\psi_4$ and $I_4$ and they do not change type under time reversal, using the first rule we find that $I_4$ transforms in the same way as $\psi_4$. That is, $T^2=-1$ for the charged boson. From this, we can easily tell that $\psi_0$, the chargeless fermion, transforms as $T^2=1$. Finally, $\{\si_1,\si_7\}$ have the same $T^2=\eta=\pm 1$, while $\{\si_3,\si_5\}$ have $T^2=-\eta$.  This information is summarized in the following table of $T^2$ values:

\begin{table}
\be
\begin{array}{|c|c|c|c|c|c|c|c|c|}
\hline
     & 0 & 1 & 2 & 3 & 4 & 5 & 6 & 7 \\ \hline
I    & 1 &   & \times  &   & -1 &   & \times &   \\ \hline
\si  &   & \eta &   & -\eta&   & -\eta&   & \eta \\ \hline
\psi & 1&   & \times&   & -1&   & \times&  \\ \hline
\end{array}
\label{T_Tsquared}
\ee  
\caption{Time reversal symmetry action on the T-Pfaffian. The semion-antisemion $(I_2,\,\psi_2)$  (as well as $(I_6,\,\psi_6)$) are exchanged by time reversal symmetry. For the remaining quasiparticles, Kramers degeneracy (${\mathcal T}^2=\pm1$) can be assigned as shown above. The electron $\psi_4$ is a Kramers doublet as is $I_4$. The non Abelian anyons can have two possible ${\mathcal T}^2$ assignments given by $\eta=\pm1$. }
\end{table}

{\em Two T-Pfaffians:} Hence there are two possible ways that time reversal symmetry can act on the T-Pfaffian state, labeled  with $\eta=\pm 1$. These two states can be mapped into each other by combining with the following $Z_2$ gauge theory. Consider a $Z_2$ gauge theory where the gauge charge $\rm e$ and the gauge flux $\rm m$ both transform as $T^2=-1$ and $\rm e$ carries $U(1)$ charge $-1$ (in units of the electron charge) while $\rm m$ is neutral. This particular $Z_2$ gauge theory cannot be realized in a purely 2D system as discussed in Ref.\cite{Vishwanath2013}, where it was termed the eTmT state, and realizes the surface topological order of a bosonic SPT phase protected by time reversal symmetry (charge attached to the $\rm e$ particle is an unimportant difference).  Bring such a $Z_2$ gauge theory on top of the $\eta=1$ T-Pfaffian state and condense the boson pair ${\rm e}I_4$. The combination of $\rm e$ and $I_4$ is charge neutral and transforms as $T^2=1$, therefore, the condensate preserves both symmetries. After the condensation, the gauge flux $\rm m$ gets bounded to the $\sigma$ particles in the T-Pfaffian state in order to commute with the condensate while all the abelian particles in the T-Pfaffian state remains. Therefore, the particle content in the resulting theory is the same as the original T-Pfaffian state, but the time reversal representation of the $\sigma$ particles change from $\eta=1$ to $\eta=-1$. The $U(1)$ charge carried by the particles remains the same. Thus, these two states differ by a particular bosonic topological superconductor. 


\section{Comparison with topological insulator surface state}
\label{consistency}

From the discussion in the previous sections we see that a possible definition of time reversal and charge conservation symmetry action does exist for the anyonic excitations in the T-Pfaffian state. Therefore, the T-Pfaffian state could potentially be realized on the surface of 3D systems with$\mathcal T$ and charge conservation symmetry, although  not  in purely 2D due to the chiral edge modes in T-Pfaffian. In the next section, we show that such a 3D realization indeed exists by presenting an exactly solvable model using the Walker-Wang construction\cite{Walker2012}. Due to the nontrivial symmetry action in T-Pfaffian, the 3D bulk of the system must have some nontrivial symmetry protected topological order. That is, the model is a 3D topological insulator. But is it {\em the}  topological insulator realized in free fermion systems or some other previously unknown topological insulator which is only possible in strongly interacting systems?


To answer this question definitely, we would need to find some topological invariants for different topological insulators and compute them for this system. However, we do not know how to do this. In the following, we will check certain properties of this model and see if it is consistent with what we know about the free fermion topological insulator. We find that: 1. two copies of T-Pfaffian is trivial which is consistent with the $Z_2$ classification of the free fermion topological insulator; 2. by breaking time reversal but not charge conservation symmetry, the topological order can be removed. Now, surface domain walls between regions with opposite $\mathcal T$ breaking carry gapless 1D modes with   $c_{-}=1$ and $\sigma_{xy}=1$ , which is known to happen in the free fermion topological insulator. This also implies $\theta=\pi$;  3. for one of the T-Pfaffian states ($\eta=1$ or $-1$), the topological order can be removed by breaking charge conservation but not time reversal symmetry, a property expected for free fermion topological insulator surface states. While we do not explicitly construct the route to removing topological order, we demonstrate this to be a logical consequence. 


\subsection{Two copies of T-Pfaffian is trivial}

The free fermion topological insulator (TI) has a $Z_2$ classification. That is, if we take two copies of the free fermion TI and allow interactions between them the surface state can be made trivial without breaking either time reversal or charge conservation symmetry. Therefore, if the T-Pfaffian state is realized on the surface of the free fermion TI, we should be able to take two copies of it and removed the topological order without breaking time reversal or charge conservation symmetry. This is indeed the case as we show below.

Suppose that we have two T-Pfaffian states whose anyons are labeled as $\{I_k,\si_k,\psi_k\}$ and $\{\t{I}_k,\t{\si}_k,\t{\psi}_k\}$. We can condense the following set of composite bosonic particles without breaking time reversal or charge conservation:
\be
I_2\t{\psi}_6,\psi_2\t{I}_6,I_6\t{\psi}_2,\psi_6\t{I}_2,I_4\t{I}_4,\psi_0\t{\psi}_0,\psi_4\t{\psi}_4
\ee 
Note first that each composite particle listed above has bosonic self and mutual statistics, therefore they can be condensed together. Also, each composite particle has charge $0 \text{mod}\ 8$, hence condensing them does not break charge conservation. Moreover, the composite particles either map to themselves under time reversal or appear in time reversal pairs. Finally, they all transform as $T^2=1$ under time reversal. Therefore, the condensate does not break time reversal either.

After condensing these particles, the nonabelian $\si_k$ and $\t{\si}_k$ particles are all confined. Some of the composite $\si\t{\si}$ type particles remains, which up to the condensed particles include
\be
\si_1\t{\si}_3, \si_1\t{\si}_7
\ee
The abelian particles that remain include (up to the condensed particles)
\be
I_4,\psi_0,\psi_4
\ee
In the resulting theory, the $\si\t{\si}$ particle splits into two abelian particles and the theory is equivalent to the product of a free fermion part $\{I,\psi_4\}$ and a simple $Z_2$ gauge theory part
\be
I,e,m,\epsilon
\ee
$e$ and $m$ comes from $\si_1\t{\si}_7$. They are bosons, carry charge $0$ and are invariant and transform as $T^2=1$ under time reversal. $\epsilon$ comes from $\psi_0$. It is a fermion, has charge $0$ and maps to itself and transform as $T^2=1$ under time reversal. Therefore, the total theory is trivial under time reversal and charge conservation symmetry and can be realized in 2D.

\subsection{Breaking time reversal symmetry and confinement}

To remove the topological order in the T-Pfaffian surface state by breaking time reversal but not charge conservation symmetry, we can bring a 2D fractional quantum Hall state with the T-Pfaffian topological order and couple it to the T-Pfaffian surface state. The 2D state has $c_{-}=1/2$ and $\sigma_{xy}=1/2$, therefore is not time reversal invariant. But it does have charge conservation symmetry. 

We label the anyons in the surface T-Pfaffian state as $\{I_k,\si_k,\psi_k\}$ and those in the 2D T-Pfaffian state as $\{I'_k,\si'_k,\psi'_k\}$. Condense the following composite particles:
\be
I_2\psi'_6,\psi_2I'_6,I_6\psi'_2,\psi_6I'_2,I_4I'_4,\psi_0\psi'_0,\psi_4\psi'_4
\ee 
Note that this condensation is very similar to the one we applied in the previous section to two copies of T-Pfaffian surface states. However, here the operation breaks time reversal from the beginning because the 2D T-Pfaffian state breaks time reversal symmetry. After this condensation, the surface state is reduce to the product of a charge neutral $Z_2$ gauge theory with anyons $\{I,e,m,\epsilon\}$ together with a charged fermion. We can further remove the topological order by condensing the $e$ particle in the $Z_2$ gauge theory.

To break the time reversal symmetry in the opposite way and remove the topological order, we bring a time reversed copy of the 2D T-Pfaffian state with anyons $\{\bar{I}'_k,\bar{\si}'_k,\bar{\psi}'_k\}$ and couple it in the time reversed way to the surface T-Pfaffian state. The statistics in the time reversed copy of the 2D T-Pfaffian is the complex conjugate of that in the original 2D T-Pfaffian state. Therefore, in this new combination, we would condense
\be
\psi_2\bar{\psi}'_6, I_2\bar{I}'_6,\psi_6\bar{\psi}'_2,I_6\bar{I}'_2,I_4\bar{I}'_4,\psi_0\bar{\psi}'_0,\psi_4\bar{\psi}'_4
\ee
The resulting theory is again composed of a neutral $Z_2$ gauge theory with $\{I,\bar{e},\bar{m},\bar{\epsilon}\}$ and a charged fermion. By condensing $\bar{e}$, we remove the topological order completely.

Between the 2D T-Pfaffian state and its time reversal copy, there is a $c_{-}=1$ and $\si_{xy}=1$ edge. Condensation in the system does not affect $c_{-}$ and $\si_{xy}$. Therefore, we can break time reversal symmetry in opposite ways on the surface T-Pfaffian state, remove any topological order, and be left with a $c_{-}=1$ and $\si_{xy}=1$ chiral edge between the two regions. This is what is known to happen on the free fermion TI surface starting from the gapless Dirac cone surface state.

\subsection{Breaking charge conservation symmetry}
\label{breakingT}

When the surface of the free fermion topological insulator is gapless, the surface Dirac cone can be gapped out (in a topologically trivial way) by inducing superconductivity on the surface without breaking time reversal symmetry. If the T-Pfaffian can be realized as the topologically ordered surface state of the free fermion TI, we would like to see that the topological order can be removed by condensing charge without breaking time reversal symmetry. In the T-Pfaffian state, it is not obvious how this can be achieved. For example, one might want to condense the charged boson $I_4$ and simplify the topological order. However, $I_4$ transforms under time reversal as $T^2=-1$. Therefore, condensing $I_4$ necessarily breaks time reversal symmetry. 
The other Abelian particles are not bosons and cannot be directly condensed. We show that such a removal of topological order can be achieved for one of the T-Pfaffian state ($\eta=1$ or $\eta=-1$) by combining with a 2D topological order which has time reversal but not charge conservation symmetry and then condensing composite bosonic particles. Therefore, one of the T-Pfaffian state is consistent with being the surface state of the free fermion TI. Our argument proceeds in the following steps:. 


1. The `modularized' T-Pfaffian is a time reversal symmetric bosonic topological state. As a fermionic topological order, the T-Pfaffian state has $Z_2$ fermion parity symmetry. We can gauge the $Z_2$ symmetry and obtain a `modularized' bosonic topological theory where the local fermion has a mutual $-1$ statistics with the $Z_2$ fluxes. Such a gauging process is not unique and we find that one of the possible `modularized' theories has time reversal symmetric fusion and braiding statistics. The gauging process and the resulting theory is described in detail in appendix \ref{gauge}. This step works for both versions of the T-Pfaffian ($\eta=\pm 1$). 

2. One of the `modularized' T-Pfaffians can be realized in 2D with time reversal symmetry. As a bosonic topological order with time reversal symmetric fusion and braiding statistics, the `modularized' T-Pfaffian states can be either realized in 2D time reversal symmetric systems or on the surface of 3D bosonic topological superconductors. By simply looking at the theories, it is hard to tell which is the case. However, useful information can be obtained from our knowledge of bosonic topological superconductors. We know that bosonic topological superconductors has a $Z_2\times Z_2$ classification\cite{Chen2011, Chen2012a, Vishwanath2013}. These are composed of (i) the nontrivial state in the first $Z_2$ has three fermion surface topological order (which is chiral when realized in 2D) while (ii) the nontrivial state for the second $Z_2$ has both the electric and magnetic charges transforming as ${\mathcal T}^2=-1$ and is a nonchiral surface topological order\cite{Vishwanath2013,Burnell2013,Wang2013}, labeled eTmT in Ref.\cite{Vishwanath2013,Wang2013} . Because the modularized T-Pfaffian theories are nonchiral, they must belong to either the trivial or the nontrivial case of the second $Z_2$ (i.e. eTmT surface topological order). Moreover, the two modularized T-Pfaffians differ by exactly eTmT. To see this, take the eTmT surface topological order 
-- and bring it on top of the $\eta=1$ modularized T-Pfaffian. Condense the composite particle $eI_4$. Because both $e$ and $I_4$ are both Kramer doublets, such a condensation does not violate time reversal symmetry. The resulting theory after the condensation is exactly the $\eta=-1$ modularized T-Pfaffian state. Therefore, one of the modularized theory can be realized in 2D with time reversal symmetry while the other is the surface state of a nontrivial bosonic topological superconductor (eTmT surface topological order). One last piece of missing information is which one is which. We do not know the answer to this question right now.

3. Combining the `modularized' T-Pfaffian with T-Pfaffian, the topological order can be removed without breaking time reversal symmetry. Now bring the modularized T-Pfaffian state (complex conjugated) on top of the corresponding T-Pfaffian state. Condense fermion-fermion pair to confine the $Z_2$ gauge field and we obtain a doubled T-Pfaffian theory. The topological order in this doubled theory can be completely removed by condensing pairs of corresponding particles, e.g. $I_2I'_2$, which does not violate time reversal symmetry. Therefore, for one of the T-Pfaffians, the surface topological order can be removed through combination with a 2D $\mathcal{T}$ invariant topological state and condensing bosons in a time reversal invariant way. 

Therefore, we can conclude from the previous analysis that one of the T-Pfaffian states , on breaking charge conservation symmetry, is equivalent to the superconducting surface of the free fermion topological insulator, while the other differs from it by a bosonic topological superconductor (with surface topological order eTmT). Thus one of the T-Pfaffian states passes all the physical requirements expected of TI surface topological order - Z$_2$`ness', , thermal and electrical Hall (equivalent to $\theta=\pi$) conductivity on $\mathcal T$ breaking surface domain walls, and surface superconductor  with $\mathcal T$ symmetry and free of topological order. However at this moment, we cannot tell whether this is the  $\eta=1$ state or the $\eta=-1$ T-Pfaffian state.

\subsection{Connecting Surface Topological Order to the Free fermion TI}
We mention here a simple argument that allows us to connect one of the T-Pfaffian states with the free fermion TI, based on the classification result of \onlinecite{Wang2013b}. We note however that the argument above did not utilize this result. In Ref.\onlinecite{Wang2013b}, $U(1)\rtimes Z_2^T$ fermionic topological insulators in 3D were proposed to have a $Z_2^3$ classification - one $Z_2$ corresponds to the free fermion topological insulator while the remaining $Z_2^2$ refers to  neutral bosonic SPTs with time reversal symmetry. The ambiguity here is whether the T-Pfaffian(s) describe a mixture of the free fermionic topological insulator with 3D bosonic SPT phases. We can prove that there is a T-Pfaffian state which is to be identified with the free fermion TI, with no bosonic SPT mixture assuming the classification result above.  

First let us discuss possible Bosonic SPT phases with $U(1) \rtimes Z_2^T$ symmetry which have a $Z_2^3$ classification. The root states in terms of surface topological order are (i) three fermion state (ii) the toric code with $e$ and $m$ transforming as $T^2=-1$ (eTmT state) and (iii) the toric code where both $e$ and $m$ are charge $1/2$ of the Cooper pair. However, in the presence of electrons the classification is reduced to $Z_2^2$ for these bosonic SPTs. One can combine say the electron with (iii) to obtain a mixture of (i) and (ii). Therefore one can take the two neutral states (i) and (ii) as the relevant topological orders \cite{Wang2013b}.

The remaining question is - is the T-Pfaffian a mixture of fermionic TI and one or both of the bosonic SPTs (i), (ii)? Note, if it is a mixture, then breaking charge conservation symmetry is not sufficient to eliminate the topological order.  The T-Pfaffian cannot contain the state (i) since T-Pfaffian has $\kappa_{xy}=\frac12 \kappa_0$ and $\sigma_{xy}=\frac12 \sigma_0$ in one realization while the mixture with state (i) would have an additional thermal Hall conductivity of $\pm 4\kappa_0$, while retaining the same charge Hall conductivity. Here $\sigma_0=e^2/h$ and $\kappa_0= L_0 \sigma_0 T$ where $L_0 = \frac{\pi^2}3\left ( \frac{k_B}{e} \right )^2$ is the Lorentz number and $T$ is the temperature.

Therefore the T-Pfaffian can at best differ from the free fermion TI by the eTmT state (ii). However, the two versions of the T-Pfaffian differ from one another by exactly this state, which has a $Z_2$ classification. Hence one of the two T-Pfaffian states represents the surface of the free fermion TI.

\section{Walker Wang Construction and More on Time Reversal Symmetry}
\label{WW}

In this section, we construct the 3D model with time reversal symmetry which realizes the T-Pfaffian theory on its surface and have time reversal acting in the way expected. In section \ref{WWa}, we introduce the basic idea of the Walker Wang construction and explain in section \ref{WWT} how it allows us to determine the local time reversal symmetry transformation ($T^2=\pm 1$) for the anyons. 
We try to first present the basic picture and the general idea underlying the Walker Wang construction in this section without going into too much details, which is saved for section \ref{WWd} where we give the exactly solvable Hamiltonian and address some related subtleties.

\subsection{Walker-Wang: general idea}
\label{WWa}

The Walker-Wang construction provides a way to write down an exactly solvable 3D model which realizes a particular topological order on the 2D surface of the system\cite{Walker2012}. Given all the fusion and braiding data of a 2D anyon theory, the Walker-Wang prescription gives the local Hilbert space, terms in the Hamiltonian and ground state wave function of a 3D model such that the 2D anyon theory emerges on the surface of the system. While it is not surprising that 2D anyon theories can be realized on the surface of 3D systems, the Walker-Wang construction is useful in the following ways: (1) it provides exactly solvable 3D models to realize `chiral' 2D topological orders, for which a 2D exactly solvable model is not known to exist; (2) the 3D Walker-Wang model can have extra symmetry than is possible on the topological order in a purely 2D system. That is, the surface of the Walker-Wang model can realize symmetry enriched topological orders that is not possible in 2D, which is a result of the nontrivial symmetry protected topological order in the 3D bulk of the system. In our previous works, we have explored this property of the Walker-Wang model in the case of bosonic and fermionic topological superconductors, demonstrating the existence of time reversal invariant topological orders which are impossible in purely 2D systems but can be realized on 3D surface. Here, we use a similar strategy to study fermionic topological insulators and show that the T-Pfaffian state can be realized on the surface of a 3D system with time reversal and charge conservation symmetry while it is not possible in 2D with the same symmetry. 

In this section, we are not going to explain all the details related to the exactly solvable Hamiltonian, but only focus on the basic idea of the Walker-Wang construction and show how it allows us to determine that $T^2=-1$ for the electrons. ($T^2$ for other anyons can also be determined.) 

The basic idea underlying the Walker-Wang construction is very intuitive. The model is constructed such that the ground state wave function is a superposition of 3D loops (more precisely `string nets', in the sense of \cite{Levin2005}) labeled by the anyon types, which describes the 2+1D space time trajectory of the anyons. The amplitude for a given configuration of these loops $C$ in the 3+1D wave function $\Psi_{\rm 3D}(C)$ is determined by the expectation value of the corresponding Wilson loop operators in the 2+1D TQFT (Topological Quantum Field Theory) which describes how the anyon world lines twist and intertwine with each other; i.e. we have:
\be
\Psi_{\rm 3D}(C) = \langle W(C)\rangle_{\rm 2+1TQFT}
\label{ground_state}
\ee
This is similar in spirit to eg. Quantum Hall wave functions, which are related to the space-time correlations of their edge states. Here, since we demand a topologically ordered boundary state, the expectation values are taken in the boundary TQFT. 

The wave function for the T-Pfaffian Walker-Wang model hence contains 12 different string types corresponding to the 12 different anyons in the theory which can braid and fuse according to the fusion rules of T-Pfaffian. The strings have directions. If the direction of a string related to anyon $i$ is reversed, it becomes a string related to the anti-particle, $i^*$. Since the twisting and intertwining of the anyon world lines may depend on the angle of view, in order to calculate the amplitude of the string-net configurations, we need to pick a particular projection of the 3D loops onto a 2D surface. The projection we will use is also shown in Fig.\ref{trivalent_lattice}. 

Having fixed a projection, the amplitude of each configuration can be obtained using the braiding and fusion rules given by the $R$ and $F$ matrix of the T-Pfaffian theory, which is the product of the $R$ and $F$ matrices of the Ising part and the $U(1)_8$ part.
The $R$ matrix for the Ising part reads
\be
\begin{array}{l}
R^{I,*}_{*}=R^{*,I}_{*}=1, R^{\si,\si}_{I}=e^{i\frac{\pi}{8}}, R^{\si,\si}_{\psi}=e^{-i\frac{3\pi}{8}} \\
R^{\si,\psi}_{\si}=R^{\psi,\si}_{\si}=i, R^{\psi,\psi}_{I}=-1
\end{array}
\ee

The $R$ matrix in the $U(1)$ part is 
\be
R^{k_1,k_2}_{(k_1+k_2)\text{mod}\ 8}=e^{i\frac{2\pi}{16}k_1k_2}
\ee

The $F$ matrix for the Ising part is
\be
\begin{array}{l}
\left[F^{\si,\si,\si}_{\si}\right]_{\al,\bt}=\frac{1}{\sqrt{2}}\begin{pmatrix} 1 & 1 \\ 1 & -1 \end{pmatrix}  \\
\left[F^{\si,\psi,\si}_{\psi}\right]_{\si,\si}=\left[F^{\psi,\si,\psi}_{\si}\right]_{\si,\si}=-1
\end{array}
\ee
where $\al, \bt = I, \psi$. All other terms being $1$.

The $F$ matrix for the $U(1)$ part is 
\be
\begin{array}{l}
\left[F^{k_1,k_2,k_3}_{(k_1+k_2+k_3)\text{mod}\ 8)}\right]_{(k_1+k_2)\text{mod}\ 8,(k_2+k_3)\text{mod}\ 8} \\ \nonumber
=e^{i\frac{\pi}{8}k_1(k_2+k_3-(k_2+k_3)\text{mod}\ 8)}
\end{array}
\ee
It can only take value $\pm 1$.

\begin{figure}[htbp]
\includegraphics[width=0.8\linewidth]{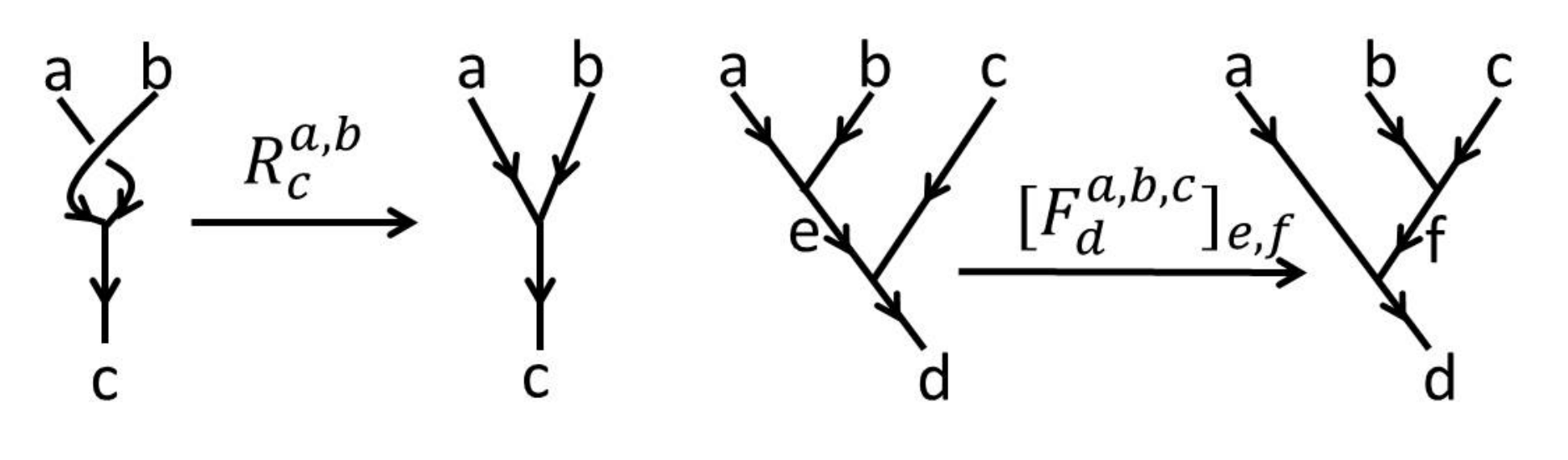}
\caption{(color online) Braiding and fusion moves on the string-net configurations.}
\label{RF}
\end{figure}

Using the braiding and fusion moves as illustrated in Fig.\ref{RF}, we can deform any string-net configurations configuration to a set of isolated loops. The change in the amplitude of the configurations is given by the $F$ and $R$ matrices. The isolated loops can be removed with the change in amplitude by
\be
\begin{array}{l}
\Delta_{I_k}=\Delta_{\psi_k}=1, \ k=0,2,4,6 \\ 
\Delta_{\si_k}=-\sqrt{2}, \ k=1,3,5,7
\end{array}
\ee
Using these set of rules, the amplitude of any string-net configuration can be determined (relative to the all $I_0$ configuration).

\begin{figure}[htbp]\vspace{-0pt}
\includegraphics[width=0.6\linewidth]{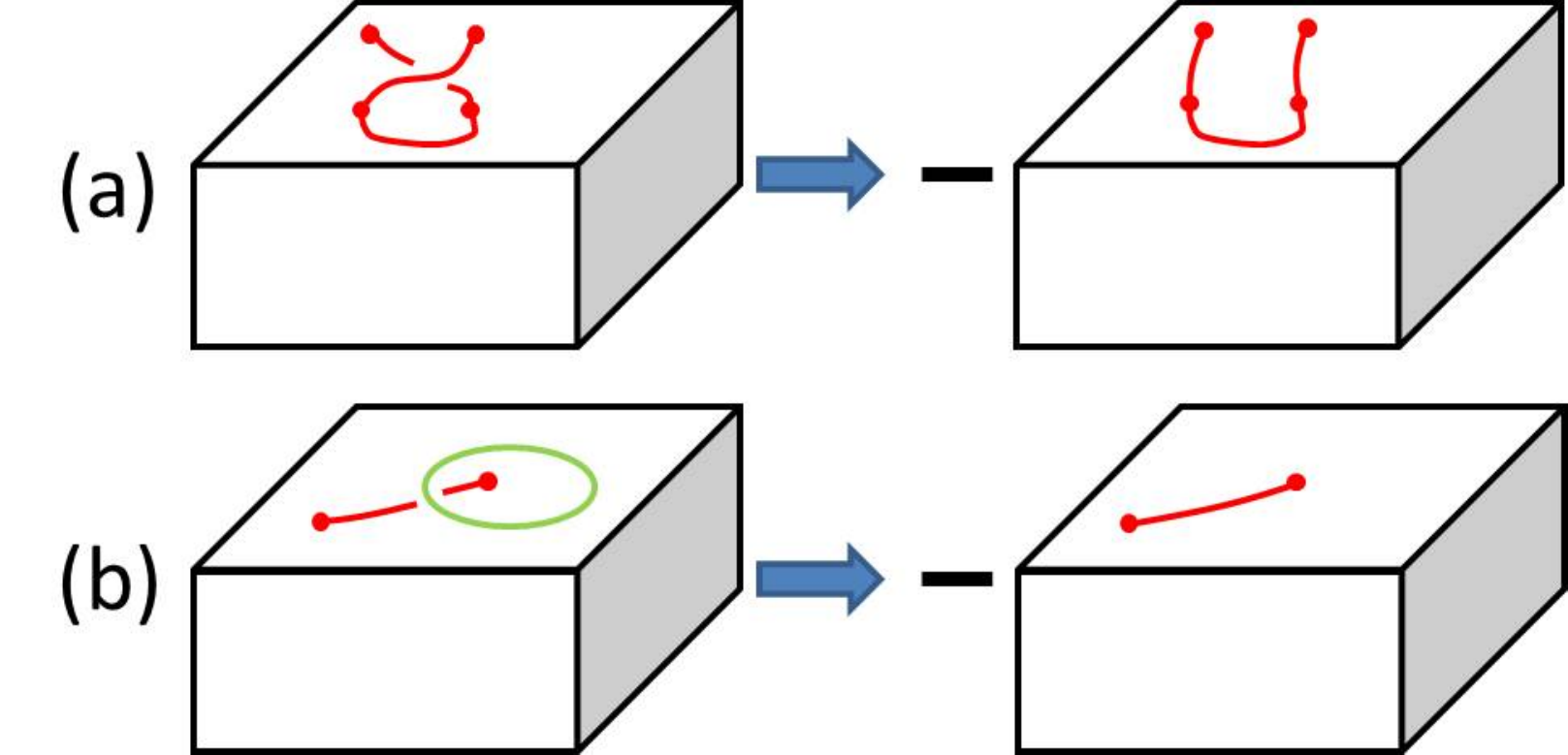}
\caption{(color online) The anyonic excitations on the surface are created by open strings. The end of strings of type $\al$ are anyonic excitations of type $\al$ with the expected statistics. This can be seen from the braiding statistics of the strings generating (a) the exchange and (b) the braiding of the end of strings.}
\label{surface_anyon}
\end{figure}

Such a bulk wave function encodes the statistics on the surface, as we show below. Anyonic excitations can be created by adding open strings to the surface. The wave function becomes a superposition of all string-net configurations in which the corresponding strings end at the positions of the excitations. Then we can check the statistics of the excitations by tracking these open strings. Suppose we exchange two string ends of the same type $\al$ (as show in Fig.\ref{surface_anyon} (a))  by crossing two red string segments on the surface. The two $\al$ anyons fuse to a $\bt$ anyon. (Fig.\ref{surface_anyon}(a) shows one possible string-net configuration.)  This twist in the string-net configuration (relative to the string-net configuration before exchange) can be removed to bring the strings back to their original form, but this results in a factor of $R^{\al,\al}_{\bt}$. Therefore, exchanging end of strings of the same type adds a $R^{\al,\al}_{\bt}$ factor to the total wave function, which is equivalent to saying that the ends of the strings are anyons with self statistics given by $R^{\al,\al}_{\bt}$. Similarly one can check, with linked loops on the surface as shown in Fig.\ref{surface_anyon}, that string ends of different types have mutual statistics given by the corresponding $R$ matrix element.

\begin{figure}[htbp]\vspace{-0pt}
\includegraphics[width=0.6\linewidth]{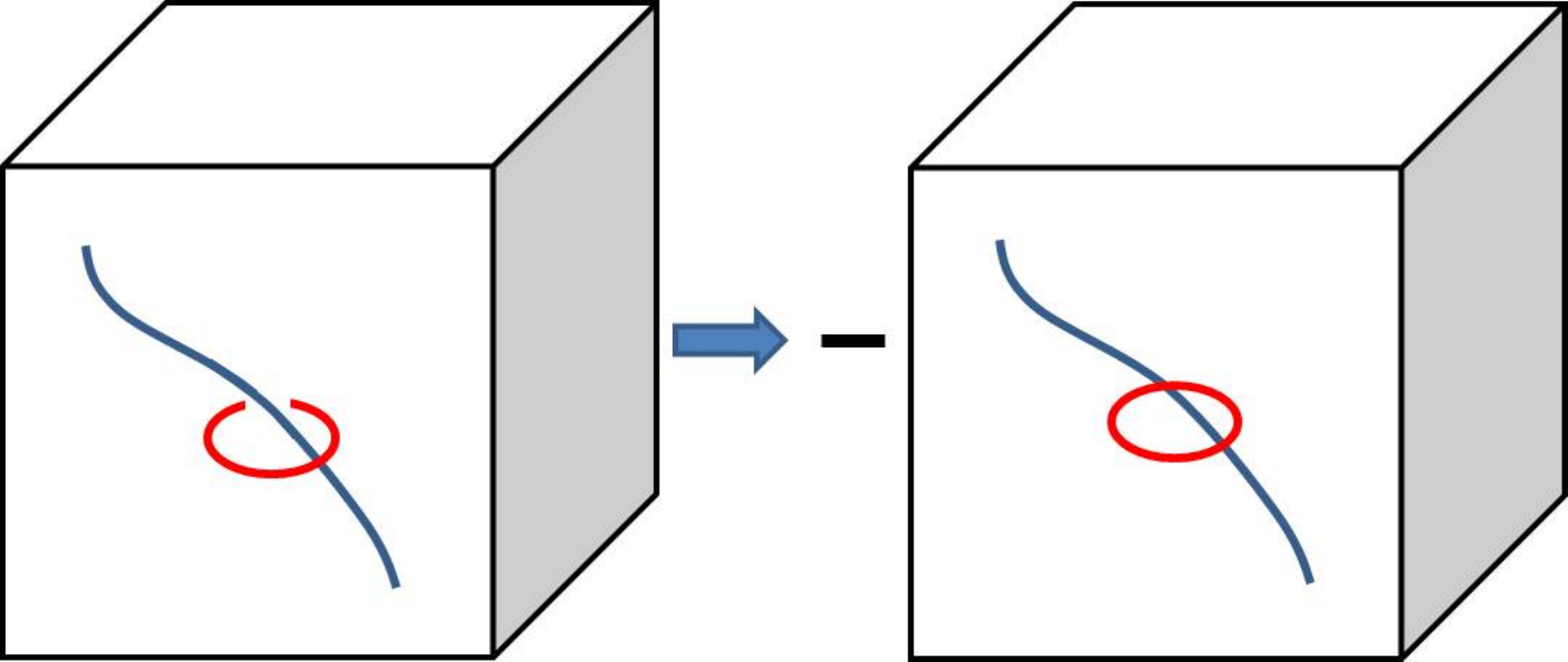}
\caption{(color online) If an anyon $\al$ has nontrivial braiding with some other anyon $\bt$, then the open string of type $\al$ in the bulk can change the quantum fluctuation phase factors of $\bt$ loops along its length, which costs finite energy. Therefore, the end of strings of type $\al$ are confined in the bulk. Otherwise, the end of string of type $\al$ is a deconfined point excitation, which must be either bosonic or fermionic in a 3D system.}
\label{confinement}
\end{figure}

Open strings in the bulk also create excitations in the ground state. However, if the corresponding anyon has nontrivial braiding with any other anyon, the excitation energy grows linearly with the string length, leading to confinement of the particles at the ends of the strings.
To see the confinement, consider an open string of type $\al$ (colored blue) in the bulk which is circled by a small ring of a different string type $\bt$ (colored red), as shown in Fig.\ref{confinement}
Using the braiding rule between the loops we find that the linking between the ring and the open string can be removed together with a phase factor 
\be
S_{\al,\bt}=\frac{1}{D}\sum_{\ga} d_{\ga} R^{\bar{\bt}\al}_{\ga}R^{\al\bar{\bt}}_{\ga}
\ee
where $\ga$ is the fusion product of $\al$ and $\bt$, $d_{\ga}$ is its quantum dimension and $D=\sqrt{\sum_{\al}d_{\al}}$.
If $S_{\al,\bt} \neq 1$, the open string changes the quantum fluctuation phase factors of small loops along its length, which costs finite energy. Therefore, the string's endpoints cannot be separated very far, and the corresponding anyonic excitations in the bulk are confined. If $S_{\al,\bt}=1$ for all $\bt$, then the end of string of type $\al$ can be a deconfined point excitation in the 3D bulk of the system, which has to be either a boson or a fermion.

In the T-Pfaffian state, all anyons have nontrivial braiding with some other anyon except the electron $I_4$. The electron is a local excitation in the 2D topological order and hence has trivial braiding with any other anyon. Therefore, in the T-Pfaffian Walker-Wang model, the only deconfined excitation in the 3D bulk is the electron. However, open strings lying on the surface, where no loops can encircle them, give rise to deconfined excitations\cite{Keyserlingk2013}. Therefore, the 3D Walker-Wang model written in terms of the fusion and braiding rules of the T-Pfaffian theory has a deconfined electron in the bulk and deconfined quasiparticles corresponding to all the anyons in the T-Pfaffian theory on the surface. We will postpone describing all details of the exactly solvable model to section \ref{WWd}. First, let me see how the Walker-Wang construction tells us more about time reversal symmetry action in the T-Pfaffian state.

\subsection{Time Reversal Symmetry of the Walker-Wang Model}
\label{WWT}

The Walker-Wang model provides us with not only an exactly solvable model to realized the T-Pfaffian surface state, but also a more concrete setup to study the time reversal symmetry in the system. 

Because the ground state wave function is determined by the $F$ and $R$ matrices of the T-Pfaffian state, in order for the wave function to be time reversal symmetric, the time reversal symmetry action must leave the $F$ and $R$ matrices invariant. However, a quick check shows that the $F$ and $R$ matrices are not invariant under the exchange of $I_2 \leftrightarrow \psi_2$, $I_6 \leftrightarrow \psi_6$ and complex conjugation. For example, \footnote{In fact, there is no gauge choice of $F$ and $R$ such that simple exchange and complex conjugation leaves them invariant.} 
\be
\left(R^{\psi_2I_2}_{\psi_4}\right)^*=-R^{I_2\psi_2}_{\psi_4}
\ee
In order to fix this, we need to introduce extra phase factors $\alpha^{ij}_{k}$ in the time reversal symmetry action to the vertices where three strings meet. With proper choice of $\alpha^{ij}_{k}$, the $F$ and $R$ matrices can be invariant as
\be
\begin{array}{l}
\left[F^{ijk}_l\right]^*_{m,n}=\left[F^{\bar{i}\bar{j}\bar{k}}_{\bar{l}}\right]_{\bar{m},\bar{n}}\frac{\al^{ij}_m\al^{mk}_l}{\al^{jk}_n\al^{in}_l} \\ \nonumber
\left(R^{ij}_k\right)^*=R^{\bar{i}\bar{j}}_{\bar{k}}\frac{\al^{ij}_k}{\al^{ji}_k}
\end{array}
\label{FR_T_eq}
\ee
where $\bar{i}$ is the time reversal partner of $i$. That is, we need to introduce some extra vertex degrees of freedom at the branching point of strings, which gets a phase factor under time reversal symmetry action. With a proper choice of gauge for $F$ and $R$ (explained in detail in appendix \ref{FR_T}), a possible set of $\alpha$'s is: $\alpha=i$ for the vertices
\be
\begin{array}{l}
(\si_1,\si_5,I_2), \ (\si_3,\si_7,I_6), \ (I_2,I_2,I_4), \\
(I_2,\psi_2,\psi_4), \ (I_6,I_6,I_4), \ (I_6,\psi_6,\psi_4)
\end{array}
\label{T1}
\ee
and $\alpha=-i$ at their time reversal partners
\be
\begin{array}{l}
(\si_1,\si_5,\psi_2), \ (\si_3,\si_7,\psi_6), \ (\psi_2,\psi_2,I_4), \\
(\psi_2,I_2,\psi_4), \ (\psi_6,\psi_6,I_4), \ (\psi_6,I_6,\psi_4)
\end{array}
\label{T2}
\ee
and $\alpha=1$ for all other allowed vertices. Note that here we are labeling the strings at each vertex such that the corresponding anyons fuse to the vacuum, i.e. the direction of the strings are all pointing towards the vertex. The strings at each vertex are ordered in a clockwise way. (With this choice of $\alpha$, Eq. \ref{FR_T_eq} is satisfied, not for the $F$ and $R$ given above but with some other gauge choice of $F$ and $R$. This is explained in detail in appendix \ref{FR_T}.)

With this choice of $\alpha$, we find that $T^2=-1$ on all the vertices listed above in Eq. \ref{T1} and \ref{T2}, while $T^2=1$ on all other allowed vertices. But where are these $T^2=-1$ vertex degrees of freedom coming from? Actually they are related to the $T^2=\pm 1$ transformation law for each anyon type under time reversal, as we explain below.

\begin{figure}[htbp]\vspace{-0pt}
\includegraphics[width=0.6\linewidth]{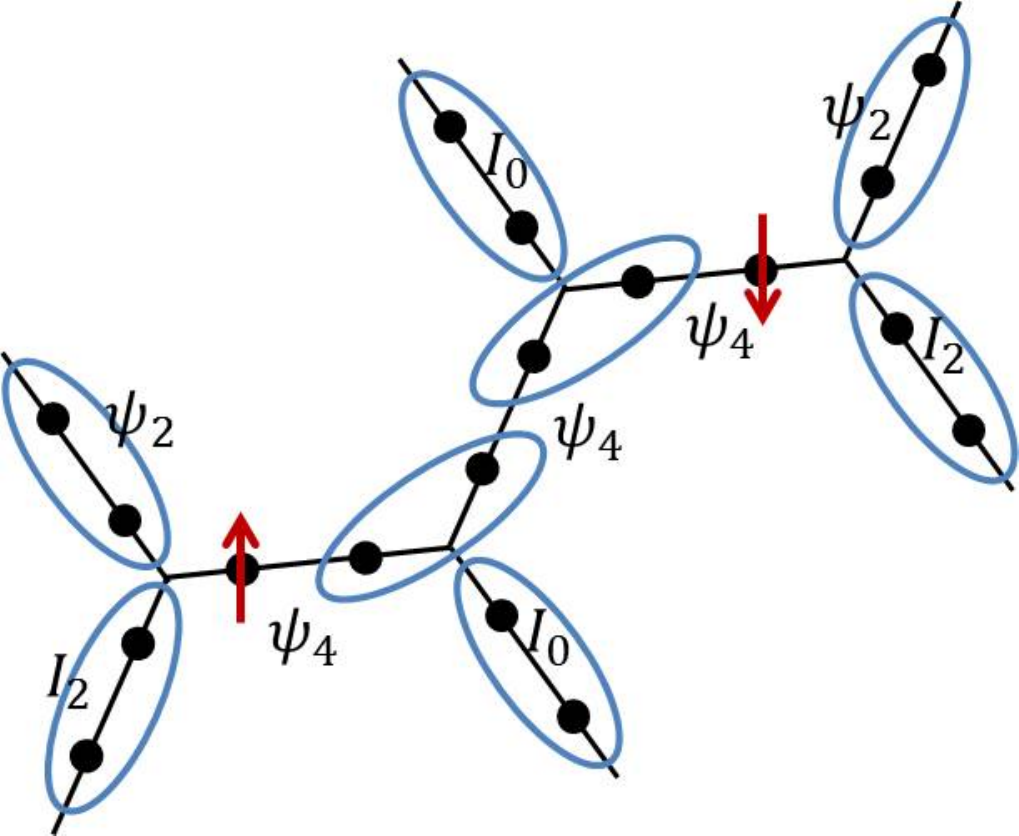}
\caption{(color online) Sample spin configuration in the decorated Walker-Wang model. Black dots represent spin $1/2$'s, and blue ellipses
represent spin singlets. The leftmost vertex $\{I_2,\psi_2,\psi_4\}$ prefers a down spin while the rightmost vertex $\{\psi_2,I_2,\psi_4\}$ prefers an up spin.}
\label{Hchain}
\end{figure}

An simple way to understand the vertex $T^2=-1$ degrees of freedom is to `attach Haldane chains' to the strings labeled by $\si_3$, $\si_5$, $I_4$ and $\psi_4$. Imagine adding pairs of spins $1/2$ to each string segment. Along the strings of types 
\be
\si_3,\ \si_5,\ I_4,\ \psi_4
\ee
the spin $1/2$'s are put into a `Haldane chain' state where spins on neighboring string segments are connected into singlet pairs, as shown in Fig.\ref{Hchain}. Along strings of all other types, the two spins on the same segment are connected into singlet pairs. A sample configuration is shown in Fig. \ref{Hchain}.

Because of the two different ways of pairing the spins, there are unpaired spins at vertices like $\{I_2,\psi_2,\psi_4\}$ and $\{\psi_2,I_2,\psi_4\}$ as shown in Fig. \ref{Hchain}. Note that because we have fixed a projection on the system, vertices $\{I_2,\psi_2,\psi_4\}$ and $\{\psi_2,I_2,\psi_4\}$ are different with different chiralities. We can choose to have the spin to point up at vertex $\{I_2,\psi_2,\psi_4\}$ and point down at vertex $\{\psi_2,I_2,\psi_4\}$. In this way, it becomes natural that time reversal adds an extra $\alpha=i$ at vertex $\{I_2,\psi_2,\psi_4\}$ and an extra $\alpha=-i$ at vertex $\{\psi_2,I_2,\psi_4\}$. In general, an unpaired spin always appears at vertices listed in Eq. \ref{T1} and \ref{T2}. We choose to have the spin to point up at vertices in Eq. \ref{T1} and to point down at vertices in Eq. \ref{T2}. Therefore, all these vertices transform under time reversal as $T^2=-1$. There are other possible ways to `attach Haldane chains' to realize the desired time reversal symmetry. We summarize them in appendix \ref{FR_T}, but the different choices do not affect our discussion in the following.

Such a spin configuration not only fixes the time reversal symmetry action on the ground state, but also determines the transformation law of the deconfined anyonic excitations on the surface of the system. For example, deconfined excitations of anyon type $I_4$ on the surface are created by adding open strings of type $I_4$ to the surface. At the point of excitation, the wave function contains vertices like $\{I_0,I_0,I_4\}$. Because $I_4$ carries a Haldane chain with it (while $I_0$ does not), at the vertex where the $I_4$ string ends, there is an extra spin $1/2$ degree of freedom which transform projectively under time reversal. Therefore, the anyonic excitation of type $I_4$ on the surface carries a projective representation of time reversal. If the symmetry if not broken, the excitation gives rise to at least a two-fold degeneracy locally.

\subsection{Walker-Wang construction: details}
\label{WWd}

In this section, we discuss the details about the exactly solvable model with T-Pfaffian surface state using the Walker-Wang construction. Readers not interested in the exact form of the Hamiltonian can skip this section. Our construction here follows closely the strategy outlined in \onlinecite{Walker2012,Keyserlingk2013} and we refer the reader there for further details.

\begin{figure}[htbp]
\includegraphics[width=0.5\linewidth]{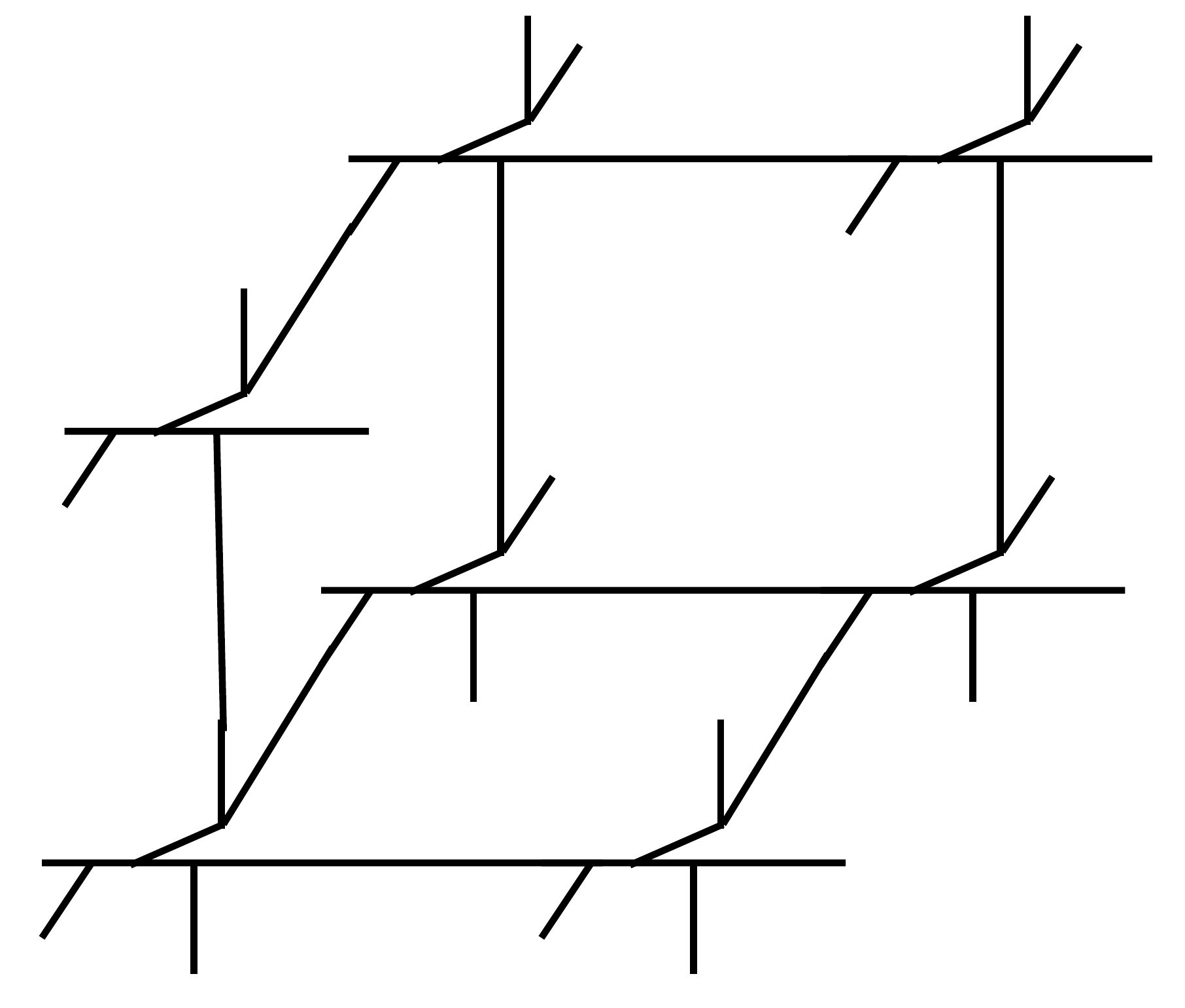}
\caption{(color online) A trivalent 3D lattice obtained by splitting the vertices in a cubic lattice (taken from \onlinecite{Walker2012}).}
\label{trivalent_lattice}
\end{figure}

The T-Pfaffian Walker-Wang model is defined on a 3D trivalent lattice. A trivalent 3D lattice can be obtained by splitting the vertices in a cubic lattice as shown in Fig.\ref{trivalent_lattice}. We have taken a particular projection of the 3D lattice onto the 2D plane. Each link in the lattice carries a $12$ dimensional spin degree of freedom, corresponding to the $12$ anyon types in T-Pfaffian. The Hamiltonian contains vertex terms $A_v$ involving all three links ending at a particular vertex and plaquette terms $B_p$ involving links around a particular plaquette
\be
H=\sum_v A_v + \sum_p B_p
\ee
The vertex term enforces the fusion rules at each vertex by giving a higher energy to all disallowed vertices where the three strings do not fuse to the vacuum. 

\begin{figure}[htbp]
\includegraphics[width=0.8\linewidth]{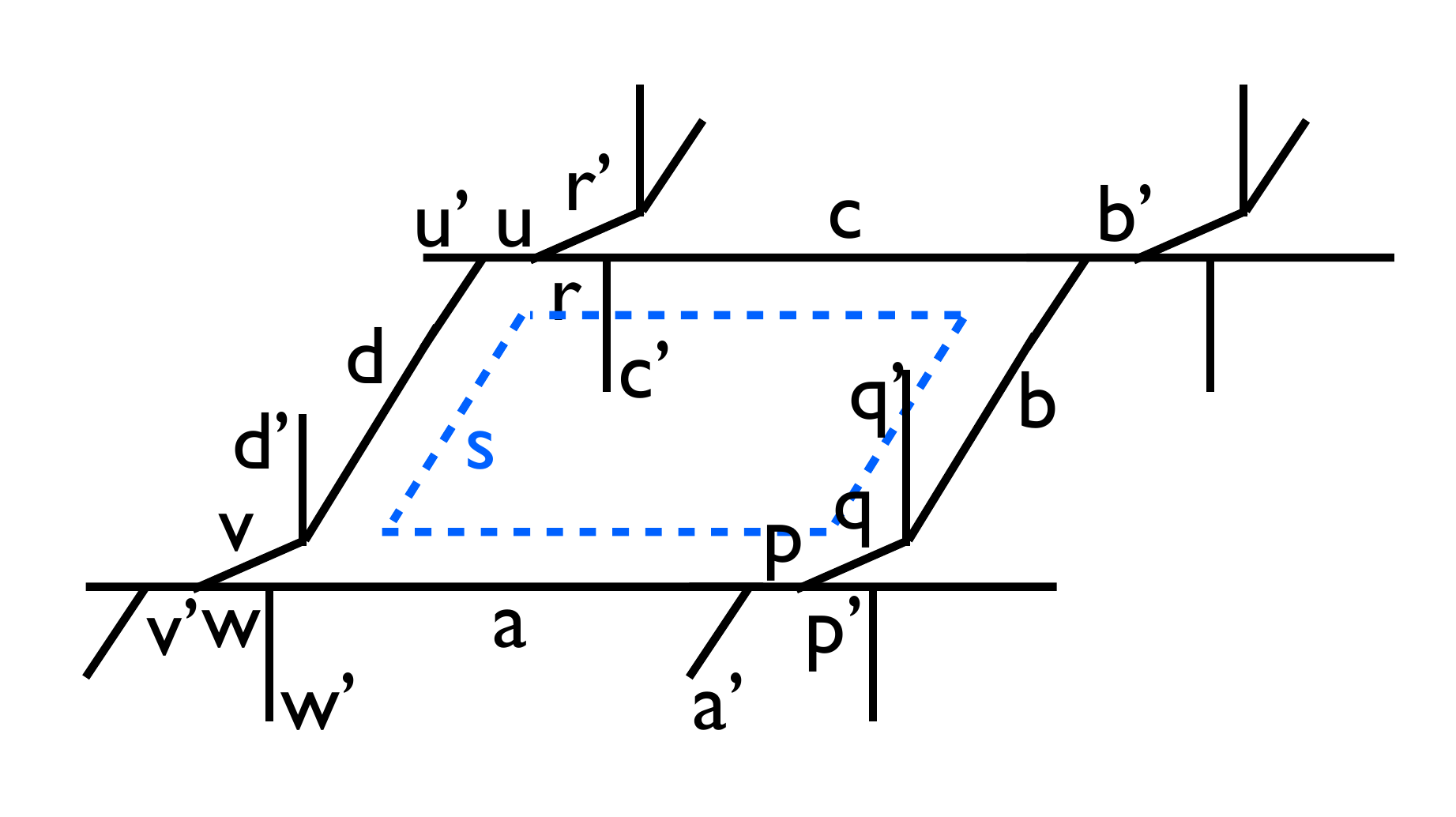}
\caption{(color online) Plaquette term in Walker-Wang Hamiltonian (taken from \onlinecite{Walker2012}).}
\label{plaquette_term}
\end{figure}

The plaquette term is a sum over terms labeled by anyon type $s$, $B_p=\sum_s d_sB^s_p$, weighted by the quantum dimension of $s$. Each $B^s_p$ can change the labels of links in a plaquette ($abcdpqruvw$), and can
also depend on the labels of adjoining links ($a'b'c'd'p'q'r'u'v'w'$) (but cannot change these). Explicitly,
the matrix element between a state with plaquette links ($abcdpqruvw$) and ($a''b''c''d''p''q''r''u''v''w''$) is
\be
\begin{array}{l}
B^{s,a,...,w}_{p,a'',...,w''}=R^{q'b}_{q}\left(R^{c'r}_c\right)^*\left(R^{q'b''}_{q''}\right)^*R^{c'r''}_{c''} \times \\ \nonumber
\left[F^{a''sp}_{a'}\right]_{ap''}\left[F^{p''sq}_{p'}\right]_{pq''}\left[F^{q''sb}_{q'}\right]_{qb''}\left[F^{b''sc}_{b'}\right]_{bc''} \times \\ \nonumber
\left[F^{c''sr}_{c'}\right]_{cr''}\left[F^{r''su}_{r'}\right]_{ru''}\left[F^{u''sd}_{u'}\right]_{ud''}\left[F^{d''sv}_{d'}\right]_{dv''} \times \\ \nonumber
\left[F^{v''sw}_{v'}\right]_{vw''}\left[F^{w''sa}_{w'}\right]_{wa''}
\end{array}
\ee

The intuition behind this complicated looking term is
that it fuses in the loop $s$ to the skeleton of the plaquette
using multiple $F$ moves, but in the process of doing
so must use $R$ moves to temporarily displace certain
links ($c'$ and $q'$ in Fig. \ref{plaquette_term}). It is possible to check that all
of these terms commute, and the the resulting ground state is a superposition of string-nets as given in Eq. \ref{ground_state}. Following the proof in \onlinecite{Walker2012,Keyserlingk2013}, we can see that the model has the bulk and surface deconfined excitations as desired. 

In order to realize time reversal symmetry on this model, we need to add two spin $1/2$ degrees of freedom to each link and put them into Haldane chains or trivial chains along different strings. At each vertex listed in Eq. \ref{T1} and \ref{T2}, there is an unpaired spin $1/2$. We add an up pointing magnetic field to this spin if it is at a vertex in Eq. \ref{T1} and a down pointing magnetic field if it is at one in Eq. \ref{T2}. The plaquette term should then be modified correspondingly to act only between the low energy configurations. It can then be checked that the Hamiltonian is indeed time reversal invariant.

 In order to add charge to this model, we can use similar tricks and attach fractional charges to the end of each link. When the links connect according to the fusion rules of the theory, the charges cancel at the vertex. When excitations are created by end of strings, extra fractional charges are present at the position of the excitation.

There is one subtle point about this construction. The Walker-Wang model as defined above is a spin model instead of a fermion model as one would expect for a 3D topological insulator. The fermions in the bulk appear as gauge fermions coupled to a $Z_2$ gauge field. Therefore, the model is topologically ordered not only on the surface but also in the bulk. Apart from the deconfined fermionic point excitations, there are also $Z_2$ flux loop excitations in the bulk around which the gauge fermions pick up a $-1$ phase factor. 

A simple way to remove the topological order in the bulk and obtain a real fermionic model is to couple the above construction with a trivial fermionic insulator and condense the fermion pairs. Because the gauge fermion transform as $T^2=-1$ under time reversal, it comes in two species $\psi_{4\uparrow}$ and $\psi_{4\downarrow}$. It is coupled to a fermionic insulator made also of $T^2=-1$ fermions $c_{\uparrow}$, $c_{\downarrow}$. One can then turn on the tunneling $c^{\dagger}_{\uparrow}\psi_{4\uparrow}+c^{\dagger}_{\downarrow}\psi_{4\downarrow}+\text{h.c.}$ and condense the particle-hole pair. The $Z_2$ flux loops are then confined because they have nontrivial statistics with the condensate. Therefore, the bulk topological order is removed and one obtains a short range entangled fermionic model as desired.


\section{The Pfaffian-antisemion state}
The process of removing topological order by breaking charge conservation but not time reversal symmetry is realized in a simpler way in the Pfaffian antisemion model \cite{Wang2013a, Metlitski2013a}. In this section we show this model can be realized as the surface state of a $\mathcal T$ and U(1) conserving bulk phase and discuss its properties compared to the free fermion TI surface.

The Pfaffian-antisemion state is a product of the anti-semion state and the Pfaffian state. We label the anyons as $\{I_k,I_ks,\si_k,\si_ks,\psi_k,\psi_ks\}$. The topological spins for the anyons are
\be
\begin{array}{|c|c|c|c|c|c|c|c|c|}
\hline
     & 0 & 1 & 2 & 3 & 4 & 5 & 6 & 7 \\ \hline
I    & 1 &   & i &   & 1 &   & i &   \\ \hline
Is    & -i &   & 1 &   & -i &   & 1 &   \\ \hline
\si  &   & e^{i\pi/4} &   & -e^{i\pi/4}&   & -e^{i\pi/4}&   & e^{i\pi/4} \\ \hline
\si s  &   & e^{-i\pi/4} &   & -e^{-i\pi/4} &   & -e^{-i\pi/4} &   & e^{-i\pi/4} \\ \hline
\psi & -1&   & -i&   & -1&   & -i&  \\ \hline
\psi s & i &   & -1 &   & i &   & -1&  \\ \hline
\end{array}
\label{TS_spin}
\ee

The Ising and anti-semion parts are both neutral while the $U(1)$ part has charge $k\frac{e}{4}$. Time reversal symmetry exchanges the following pairs of anyons $I_0s\ \& \ \psi_0s$, $I_2 \ \& \ \psi_2$, $I_4s\ \& \ \psi_4 s$, $I_6\ \& \ \psi_6$, $\si_k\ \& \ \si_ks$. Moreover, using the rules in section \ref{local_T}, we can determine the local time reversal symmetry action on anyons which do not change type under time reversal: $I_2s$, $I_4$, $I_6s$, $\psi_0$, $\psi_2s$, $\psi_4$, $\psi_6s$.

Because $I_2$ and $\psi_2$ map into each other under time reversal and braid with a $-1$, $T^2=-1$ for their fusion product $\psi_4$. Therefore, the Pfaffian-antisemion state is the surface state of a $T^2=-1$ topological insulator. Similarly, $\si_1$ and $\si_1s$ are time reversal partners. In fusion channel $I_2s$, they have trivial mutual statistics, therefore, $T^2=1$ for $I_2s$ (also for $I_6s$). In fusion channel $\psi_2s$, they have mutual semionic statistics, therefore, $T^2=-1$ for $\psi_2s$ (also for $\psi_6s$). Moreover, $I_0s$ and $\psi_0s$ are time reversal partners and have mutual semionic statistics. Therefore, $T^2=-1$ for their fusion product $\psi_0$. Finally, from the fusion of $\psi_0$ and $\psi_4$ into $I_4$ we find that $T^2=1$ for $I_4$.

 A Walker-Wang model can be written down explicitly for the Pfaffian-antisemion state as well. The model is defined on a 3D trivalent lattice similar to the T-Pfaffian case while the dimension of the Hilbert space on each link is doubled. Basis states on each link correspond to the 24 quasiparticles in the theory and the vertex term and plaquette term enforces the fusion and braiding rules. On the surface of the system, there are deconfined excitations corresponding to all 24 quasiparticles in the theory while in the bulk only the $\psi_4$ fermion is deconfined. Time reversal symmetry acts as complex conjugation, permutation of link basis states, and phase factors at each vertex. The phase factors involved are such that vertices like $\{\sigma_1,\sigma_1s,\psi_2s\}$ transform as $T^2=-1$. Such vertex transformation laws can be understood as coming from the Kramer degeneracy of $\psi_0$, $\psi_4$, $\psi_2s$, and $\psi_6s$.

\subsection{Breaking Charge Conservation, Z$_2$ ness and Breaking $\mathcal T$} A property of this model, which is different from T-Pfaffian, is that if we condense $I_2s$, the topological order can be removed without breaking time reversal symmetry. One may also observe that the fundamental $hc/2e$ vortices carry Majorana zero modes. To see this first notice that naively, a charge $e/2$ condensate would entail vorticity in multiples of $2hc/e$. A vortex of strength $hc/2e$ would induce a Berry phase of $\pi/2$ on the condensed particles. However, this can be rectified by attaching a $\sigma$ ($\sigma s$) particle to the vortex (to the anti vortex) , which has the opposite braiding statistics  That is, we can obtain the superconducting surface state of the free fermion TI starting from the Pfaffian-antisemion state.

Now we can take two copies of this state and see if it can be made trivial. We label the anyons in the two copies as $\{I_k,I_ks,\si_k,\si_ks,\psi_k,\psi_ks\}$ and $\{\t{I}_k,\t{I}_k\t{s},\t{\si}_k,\t{\si}_k\t{s},\t{\psi}_k,\t{\psi}_k\t{s}\}$. First we set aside the anti-semion parts of the theory $\{I,s\}$ and $\{I,\t{s}\}$ and analyze the two Pfaffian parts. We can again condense the following set of composite bosonic particles without breaking time reversal or charge conservation:
\be
I_2\t{\psi}_6,\psi_2\t{I}_6,I_6\t{\psi}_2,\psi_6\t{I}_2,I_4\t{I}_4,\psi_0\t{\psi}_0,\psi_4\t{\psi}_4
\ee
The particles that remain in the two Pfaffian states include (up to the condensed particles)
\be
\si_1\t{\si}_3, \si_1\t{\si}_7, I_4,\psi_0,\psi_4
\ee
This is very similar to the T-Pfaffian case except that the $\si_1\t{\si}_3$ particle has topological spin $-i$ and the $\si_1\t{\si}_7$ particle has topological spin $i$. In the resulting theory, these quantum dimension two particles split into two abelian particles $s_1$, $s_2$ (semions), which fuse to $\psi_0$.
\be
s_1 \times s_2 = \psi_0
\ee
Combined with the anti-semion part of the theory, 
the total theory after condensation is a product of two semion theories, two anti-semion theories and a trivial electron.
\be
\{I,s_1\} \times \{I,s_2\} \times \{I,s\} \times \{I,\t{s}\} \times \{I,\psi_4\}
\ee 
The semion and anti-semions parts are all neutral. Only the electron is charged. We can further condense $s_1s$ and its time reversal partner $s_2\t{s}$ without breaking any symmetry, which confines everything except the electron $\psi_4$. Therefore, the Pfaffian-antisemion state likely corresponds to a $Z_2$ topological insulator.

In order to check what happens between time reversal symmetry breaking domains, we can couple the the Pfaffian-antisemion surface state to 2D realizations of the Pfaffian-antisemion state, which must break time reversal symmetry. Following similar analysis as above, we find that if we couple a 2D Pfaffian-antisemion state on the left half of the plane, we can remove all topological order. Similarly, if we couple a time reversed 2D Pfaffian-antisemion state on the right half of the plane, with time reversed coupling, we can also removed all topological order. Between these two domains, a chiral edge state with $c_-=1$, $\si_{xy}=1$ is left behind, which is expected for the free fermion topological insulator surface state.

\subsection{Connecting the Pfaffian antisemion state to the T-Pfaffian Topological Order} 
 We can directly show that one of the two T-Pfaffians can be connected to the Pfaffian-antisemion state,i.e. is surface equivalent to it. Again, the procedure is not constructive - so we don't know which of the two it is. We do not make any assumption about the classification of interacting topological insulators (Ref. \onlinecite{Wang2013b}) but utilize a related construction.

Consider combining the T-Pfaffian state with Pfaffian-antisemion state and attempting to remove all the topological order while preserving both $T$ and charge $U(1)$ symmetry. To help this process we introduce a Z$_8$ gauge theory topological order of Cooper pairs, where the gauge charges $p_m$ $m=0,\,1,\dots,\,7$ carry charges $q_m=m\frac{2e}8$, and the gauge fluxes $v_n$, where  $n=0,\,1,\dots,\,7$ are exchanged under time reversal symmetry $n\rightarrow (8-n) \text{mod} 8$ under $T$. $v_4$ maps to itself under time reversal and we choose it to transform as a Kramer doublet. Consider condensing $I_2\tilde{\psi}_6v_2$ and its time reversed conjugate $\psi_2\tilde{I}_6v_6$. Note, here the first particle is in the Pfaffian-antisemion theory while the second is in the T-Pfaffian (with tilde). These condensates are self and mutual bosons with charge 0 and preserve time reversal symmetry. The last is particularly crucial since in the absence of the $v_2$ factors, the square of these condensates $I_4\tilde{I}_4$ would break $T$ symmetry since one of these bosons is a Kramers doublet. However here it appears as $I_4\tilde{I}_4v_4$ which is a total time reversal singlet and can be condensed. As $I_2\tilde{\psi}_6v_2$ is a strength-$2$ flux of the whole theory, all the surviving excitations after this condensation have integer electric charge in units of $e$. We can separate all the quasiparticles after this condensate into two sets, a charge neutral set and a charge $e$ set which are related to each other through combination with an electron. The neutral set form a closed modular topological theory $X$ which is $T$ symmetric (note, $T$ symmetry does not interchange particles differing by an electron since they carry different charges)\cite{Wang2013b} and the whole theory can be written as $X\times \{1,\,f\}$, where the electric charge is only carried by the electron $f$. Therefore, the T-Pfaffian state and the Pfaffian-antisemion state is surface equivalent up to either the surface of a 3D bosonic SPT of neutral boson protected by $Z_2^T$ or a 2D $T$  symmetric topological order. Again, one can exclude the 3-fermion state since the combined theory has no nontrivial edge states even when realized in 2D. So the question is whether they differ by a bosonic SPT which is eTmT (both e, m transforming as Kramers doublets). Now, since the two T-Pfaffians differ by precisely this theory, one of them is equivalent to the Pfaffian-antisemion state.

\section{Conclusions}
We have discussed the possibility of a 3D Topological Insulator with a symmetric gapped surface state and non-Abelian surface topological order. We constructed a model for a 3D Topological Insulator, with magnetoelectric response $\theta=\pi$, with a surface state given by T-Pfaffian topological order. We find that the symmetry transformation on the T-Pfaffian state can take two different forms. One of them is consistent with being the surface state of the free fermion TI while the other differ from the free fermion TI by a neutral bosonic topological superconductor. One remaining question is which specific T-Pfaffian state is connected to the free fermion surface state. We cannot yet answer this question due to the lack of a simple way to smoothly connect the T-Pfaffian state to the superconducting surface state of the free fermion TI, and is left for future work. We also constructed an exactly soluble 3D lattice model for  a somewhat more complicated topological order, the Pfaffian-antisemion theory\cite{Wang2013a, Metlitski2013a} with twice as many particles, which can be smoothly connected to the superconducting surface state. These finds are consistent with the classification result obtained in \onlinecite{Wang2013b}.


\acknowledgments

XC is supported by the Miller Institute for Basic Research in Science at UC Berkeley. XC thanks discussions with P. Bonderson, X.L. Qi, M. Fisher, T. Senthil, D. Potter, C. Wang, F. Burnell, Z.H. Wang. AV thanks Ehud Altman, Parsa Bonderson and particularly T. Senthil for several insightful discussions, and P. Bonderson, Max Metlitski and T. Senthil for discussing their results prior to publication. AV is supported by ARO MURI Grant W911-NF-12-0461.


\appendix

\section{Gauging the T-Pfaffian theory}
\label{gauge}

In order to see, with broken $U(1)$ symmetry whether the T-Pfaffian theory can be realized in two dimension with time reversal symmetry, we try to gauge the $Z_2$ fermion parity symmetry in the theory and examine whether the resulting theory can be time reversal invariant or not. There are different ways (at least 16) to consistently gauge the $Z_2$ fermion parity symmetry and we need to consider them all.

One obvious way to gauge the $Z_2$ fermion parity in the T-Pfaffian theory is into the full $Ising^*\times U(1)_8$ theory. The topological spins of all $24$ particles are listed below
\be
\begin{array}{|c|c|c|c|c|c|c|c|c|}
\hline
    & 0 & 1 & 2 & 3 & 4 & 5 & 6 & 7 \\ \hline
I   & 1 & e^{i\frac{\pi}{8}} & i & -e^{i\frac{\pi}{8}}  & 1 & -e^{i\frac{\pi}{8}} & i & e^{i\frac{\pi}{8}} \\ \hline
\si & e^{-i\frac{\pi}{8}}  & 1 & e^{i\frac{3\pi}{8}}  & -1 & e^{i\frac{7\pi}{8}}  & -1 & e^{i\frac{3\pi}{8}} & 1 \\ \hline
\psi& -1 & -e^{i\frac{\pi}{8}} & -i & e^{i\frac{\pi}{8}}  & -1 & e^{i\frac{\pi}{8}} & -i & -e^{i\frac{\pi}{8}} \\ \hline
\end{array}
\ee
The braiding of the $I_1$ particle correctly measures the (fractional) $Z_2$ quantum number associated with the original anyons. Therefore, $I_1$ is the $Z_2$ flux and the full table is obtained by combining $I_1$ with all the original anyons. This theory is obviously not time reversal invariant.

There are (at least) 15 other different ways to gauge the $Z_2$ fermion parity symmetry in this theory. They can be obtained by combining one of Kitaev's 16 fold way to gauge free fermion theory together with the full $Ising^*\times U(1)_8$ theory and condense the fermion-fermion pair. For example, if we consider the $0$th one in Kitaev's 16 fold way, the Toric Code model with $I$, $\t e$, $\t m$, $\t \psi$, then the anyon types in the resulting theory comes in two sets: the set of anyons in the T-Pfaffian theory and the set of anyons in $\left(Ising^*\times U(1)_8 - \text{T-Pfaffian}\right)\times \t m$.  and topological spins in the resulting theory are (here we are still denoting the anyons with the label in $Ising^*\times U(1)_8$ but half of them are actually combined with $\t m$ in the Toric Code theory.)
\be
\begin{array}{|c|c|c|c|c|c|c|c|c|}
\hline
    & 0 & 1 & 2 & 3 & 4 & 5 & 6 & 7 \\ \hline
I   & 1 & e^{i\frac{\pi}{8}} & i & -e^{i\frac{\pi}{8}}  & 1 & -e^{i\frac{\pi}{8}} & i & e^{i\frac{\pi}{8}} \\ \hline
\si & e^{-i\frac{\pi}{8}}  & 1 & e^{i\frac{3\pi}{8}}  & -1 & e^{i\frac{7\pi}{8}}  & -1 & e^{i\frac{3\pi}{8}} & 1 \\ \hline
\psi& -1 & -e^{i\frac{\pi}{8}} & -i & e^{i\frac{\pi}{8}}  & -1 & e^{i\frac{\pi}{8}} & -i & -e^{i\frac{\pi}{8}} \\ \hline
\end{array}
\ee
which is the same as the $Ising^*\times U(1)_8$ theory, as we expected.

If we consider an even number element ($\nu$-th) in the 16 fold way, with anyons $I$, $\t e$, $\t m$, $\t \psi$ and topological spins $1$, $e^{i\frac{\pi}{8}\nu}$, $e^{i\frac{\pi}{8}\nu}$, $-1$, then the anyon content in the resulting theory is similar to the previous case and the topological spins are (with slight abuse of notation for anyon label)
\be
\begin{array}{|c|c|c|c|c|}
\hline
    & 0 & 1 & 2 & 3  \\ \hline
I   & 1 & e^{i\frac{\pi}{8}(\nu+1)} & i & -e^{i\frac{\pi}{8}(\nu+1)}  \\ \hline
\si & e^{i\frac{\pi}{8}(\nu-1)}  & 1 & e^{i\frac{\pi}{8}(\nu+3)}  & -1 \\ \hline
\psi& -1 & -e^{i\frac{\pi}{8}(\nu+1)} & -i & e^{i\frac{\pi}{8}(\nu+1)}  \\ \hline
    & 4 & 5 & 6 & 7 \\ \hline
I   & 1 & -e^{i\frac{\pi}{8}(\nu+1)} & i & e^{i\frac{\pi}{8}(\nu+1)}  \\ \hline
\si & e^{i\frac{\pi}{8}(\nu+7)}  & -1 & e^{i\frac{\pi}{8}(\nu+3)} & 1 \\ \hline
\psi& -1 & e^{i\frac{\pi}{8}(\nu+1)} & -i & -e^{i\frac{\pi}{8}(\nu+1)}  \\ \hline
\end{array}
\ee
The theory cannot be time reversal invariant for any even $\nu$. 

Now let's check the case for odd $\nu$. With odd $\nu$, the gauged free fermion theory has a nonabelian $Z_2$ flux $\t \si$ with topological spin $e^{i\frac{\pi}{8}\nu}$. The total theory is obtained by combining the Ising$^*$ theory, the $U(1)_8$ theory and the $\nu$-th gauged free fermion theory and condensing $\psi\t{\psi}_4$, where the subscript $4$ denotes the $U(1)$ charge. The resulting theory contains the following anyons:
\be
\begin{array}{llll}
II_0=\psi\psi_4 & II_2=\psi\psi_6 & II_4=\psi\psi_0 & II_6=\psi\psi_2 \\ \nonumber
\psi I_0=I\psi_4 & \psi I_2=I\psi_6 & \psi I_4=I\psi_0 &  \psi I_6=I\psi_2 \\ \nonumber
\si I_1=\si \psi_5 & \si I_3=\si \psi_7 & \si I_5=\si \psi_1 & \si I_7=\si \psi_3 \\ \nonumber
I\si_1=\psi\si_5 & I\si_3=\psi\si_7 & I\si_5=\psi\si_1 & I\si_7=\psi\si_3 \\ \nonumber
\si\si_0=\si\si_4 & \si\si_2=\si\si_6
\end{array}
\ee

Let us use the shorthand notation
\be
\begin{array}{llll}
I_0 & I_2 & I_4 & I_6 \\ \nonumber
\psi_0 & \psi_2 & \psi_4 & \psi_6 \\ \nonumber
\si^a_1 & \si^a_3 & \si^a_5 & \si^a_7 \\ \nonumber
\si^b_1 & \si^b_3 & \si^b_5 & \si^b_7 \\ \nonumber
\Omega_0 & \Omega_2
\end{array}
\ee

The $I_k$ and $\psi_k$ particles have quantum dimension $1$, the $\si_k$ particles have quantum dimension $\sqrt{2}$ and the $\Omega_k$ particles have quantum dimension $2$. The $\Omega$ particles could split into two quantum dimension $1$ particles. Here
\be
\Omega_0\times \Omega_0 = I_0 + I_4 + \psi_0 + \psi_4
\ee 
$\Omega_0$ fuses with itself into four different particles, therefore $\Omega_0$ does not split. Neither does $\Omega_2$.

The topological spins are
\be
\begin{array}{llll}
1 & i & 1 & i \\ \nonumber
-1 & -i & -1 & -i \\ \nonumber
1 & -1 & -1 & 1 \\ \nonumber
e^{i\frac{\nu+1}{8}\pi} & -e^{i\frac{\nu+1}{8}\pi} & -e^{i\frac{\nu+1}{8}\pi} & e^{i\frac{\nu+1}{8}\pi} \\ \nonumber
e^{i\frac{\nu-1}{8}\pi} & e^{i\frac{\nu+3}{8}\pi}
\end{array}
\ee

When $\nu=-1$, the anyon theory could potentially be time reversal invariant with topological spins
\be
\begin{array}{llll}
1 & i & 1 & i \\ \nonumber
-1 & -i & -1 & -i \\ \nonumber
1 & -1 & -1 & 1 \\ \nonumber
1 & -1 & -1 & 1 \\ \nonumber
e^{-i\frac{\pi}{4}} & e^{i\frac{\pi}{4}}
\end{array}
\ee
The fusion and braiding all follow from the parent $Ising^*\times Ising^* \times U(1)_8$ theory.

Under time reversal $\Omega_0$ would map into $\Omega_2$ and vice verse. Moreover, time reversal maps between the following pairs of particles
\be
I_2 \leftrightarrow \psi_2, \ I_6 \leftrightarrow \psi_6, \ \si^b_1 \leftrightarrow \si^b_7, \ \si^b_3 \leftrightarrow \si^b_5, \ \Omega_0 \leftrightarrow \Omega_2
\ee

We can check that this time reversal symmetry action is consistent with the fusion rule of the theory. 

This is in contrast to the semion-fermion theory where one of the gauged theories does have time reversal pairs of topological spins but time reversal symmetry action is not consistent with the fusion rule. In that case, the gauged theory is obtained from $Ising^* \times Ising^* \times semion$ by condensing the $\psi \t \psi$ pair. The two dimensional particle $\si \t \si$ needs to split into two one dimensional particles in order to make the theory unitary which causes the inconsistency between time reversal symmetry action and the fusion rule. However, in the gauged T-Pfaffian case, the $\si\si$ particle does not split and time reversal can be consistent with fusion rule.

So now we need more powerful tools to determine whether T-Pfaffian or its gauged version can be realized in 2D with time reversal.


The gauged T-Pfaffian theory is a bosonic nonchiral theory. Assuming it has time reversal symmetry, so it should either be (1) realizable in 2D (2) realizable on the surface of a 3D bosonic topological superconductor (the $Z_2$ class within group cohomology).

Note,  $\psi_4$ is a time reversal doublet since it is composed by fusing a pair of particles $\psi_2$, $I_2$ which are mutual semions and go into each other under $\mathcal T$ symmetry. This implies $I_4$ is a Kramers doublet too, and in that case we cannot condense $I_4$. However, we can combine this theory with the changeless T-Pfaffian (i.e. having broken U(1) symmetry), and condense the product of the $\psi_4$ particles ($C=\langle \psi_4^{\rm Gauged}\psi_4\rangle \neq 0$) in the two theories. This boson can be condensed and confines the flux excitations in the gauged theory resulting in two copies of T-Pfaffian, which can be confined. Thus, the changeless T-Pfaffian can be converted into a confined Sc without topological order, by combining it with this state. Hence, assuming T symmetry of the gauged theory, the changeless T-Pfaffian is either trivial (realizable in 2D with T symmetry) or equivalent to the surface of a 3D $\mathcal T$ symmetric bosonic SPT phase.


\section{Local time reversal symmetry action from Walker-Wang construction}
\label{FR_T}

In this section, we show how to obtain the rules for determining the local action of time reversal symmetry given in section \ref{local_T} from the Walker-Wang construction. We apply the procedure to the case of T-Pfaffian and find which anyons transform as $T^2=-1$. We discuss how various gauge choices in the problem can affect the solution.

In the Walker-Wang construction, in order to have a time reversal invariant Hamiltonian and ground state, we need to find vertex phase factors $\alpha^{ij}_k$ which satisfy
\be
\begin{array}{l}
\left[F^{ijk}_l\right]^*_{m,n}=\left[F^{\bar{i}\bar{j}\bar{k}}_{\bar{l}}\right]_{\bar{m},\bar{n}}\frac{\al^{ij}_m\al^{mk}_l}{\al^{jk}_n\al^{in}_l} \\ \nonumber
\left(R^{ij}_k\right)^*=R^{\bar{i}\bar{j}}_{\bar{k}}\frac{\al^{ij}_k}{\al^{ji}_k}
\end{array}
\ee
$T^2$ at each vertex is then obtained by
\be
(T^2)^{ij}_{k}=\left(\alpha^{ij}_{k}\right)^*\alpha^{\bar{i}\bar{j}}_{\bar{k}}
\ee
which is equal to the combination of $T^2$ for $i$, $j$ and $k$. 
\be
(T^2)^{ij}_{k}=T^2_iT^2_jT^2_k
\ee
Of course, for anyon types $i$ which change under time reversal, $T^2$ is not well defined. However, we have the constraint that
\be
(T^2_i)^*=T^2_{\bar{i}}
\ee
In particular, if $i=\bar{i}$, $T^2_i=\pm 1$. 

From these equations, we can derive the rules given in section \ref{local_T}. Moreover, we shall see that the arbitrariness of $T^2$ for $i$ which is not equal to $\bar{i}$ is taken into account naturally in these equations.

When $i$, $j$ and $k$ are all invariant under time reversal and $i$, $j$ fuse into $k$,
\be
(T^2)^{ij}_{k}=\left(\alpha^{ij}_{k}\right)^*\alpha^{ij}_{k}=1
\ee
Therefore, 
\be
T^2_k=T^2_i\times T^2_j
\ee
as given in the first rule in section \ref{local_T}.

Next, when $i\neq \bar{i}$ and $i$ and $\bar{i}$ fuse into $k$ ($k=\bar{k}$)
\be
\left(R^{i\bar{i}}_k\right)^*=R^{\bar{i}i}_{k}\frac{\al^{i\bar{i}}_k}{\al^{\bar{i}i}_k}
\ee
Therefore,
\be
(T^2)^{i\bar{i}}_{k}=\left(\alpha^{i\bar{i}}_{k}\right)^*\alpha^{\bar{i}i}_{k}=R^{\bar{i}i}_{k}R^{i\bar{i}}_k=s^{i\bar{i}}_k
\ee
where $s^{i\bar{i}}_k$ is the phase factor coming from a full braid of $i$ around $\bar{i}$ in fusion channel $k$. Moreover,
\be
(T^2)^{i\bar{i}}_{k}=T^2_iT^2_{\bar{i}}T^2_k=T^2_k
\ee
Therefore,
\be
T^2_k=s^{i\bar{i}}_k
\ee
as given in the second rule in section \ref{local_T}.

In particular in the case of T-Pfaffian, solving the equations in Eq. \ref{FR_T_eq} with the $F$ and $R$ matrices given in section \ref{WWa}, we obtain a set of solutions for $\alpha^{ij}_{k}$. Pick one possible solution and we can determine the $T^2$ transformation law for each vertex $\{i,j,k\}$ from
\be
(T^2)^{ij}_{k}=\left(\alpha^{ij}_{k}\right)^*\alpha^{\bar{i}\bar{j}}_{\bar{k}}
\ee
We find that the following vertices have $T^2=-1$
\be
\begin{array}{l}
(\si_1,\si_5,I_2), (\si_1,\si_5,\psi_2), (\si_3,\si_7,I_6), (\si_3,\si_7,\psi_6), \\
(I_2,I_2,I_4), (\psi_2,\psi_2,I_4), (I_2,\psi_2,\psi_4), (\psi_2,I_2,\psi_4) \\
(I_6,I_6,I_4), (\psi_6,\psi_6,I_4), (I_6,\psi_6,\psi_4), (\psi_6,I_6,\psi_4) \\
\text{and permutations of them}
\end{array}
\ee
The $T^2$ of each vertex is determined by the $T^2$ of all three anyons involved.
\be
(T^2)^{ij}_{k}=T^2_iT^2_jT^2_k
\ee
From this, we find that the possible set of Kramer doublet anyons has two possibilities
\begin{itemize}
\item{$I_4$, $\psi_4$, $\si_3$ and $\si_5$}
\item{$I_4$, $\psi_4$, $\si_1$ and $\si_7$}
\end{itemize}
$T^2=1$ for all other anyons.
These assignments satisfy
\be
(T^2_i)^*=T^2_{\bar{i}}
\ee
In fact, this Kramer doublet assignment can be simply determined from the rules given in \ref{local_T}.

The $F$ and $R$ matrices can change by gauge $\beta^{ij}_k$ as
\be
\begin{array}{l}
\left[F^{ijk}_l\right]_{m,n} \to \left[F^{ijk}_{l}\right]_{m,n}\frac{\beta^{ij}_m\beta^{mk}_l}{\beta^{jk}_n\beta^{in}_l} \\ \nonumber
\left(R^{ij}_k\right) \to R^{ij}_{k}\frac{\beta^{ij}_k}{\beta^{ji}_k}
\end{array}
\ee
Under this change, the $\alpha$'s change by
\be
\alpha^{ij}_{k} \to \alpha^{ij}\beta^{ij}_{k}\beta^{\bar{i}\bar{j}}_{\bar{k}}
\ee
This does not affect the value of $(T^2)^{ij}_{k}$ for each vertex but can change the value of $\alpha^{ij}_{k}$ to be anything consistent with the $(T^2)^{ij}_{k}$ value. Using this degree of freedom, we can make $\alpha$'s to be $i$ for vertices in Eq. \ref{T1}, $-i$ for vertices in Eq. \ref{T2} and $1$ for all other vertices.


The $\alpha$'s satisfying Eq. \ref{FR_T_eq} are not unique. In particular, if they change by
\be
\t{\alpha}^{ij}_{k}=\alpha^{ij}_{k}\frac{a_ia_j}{a_k}
\ee 
then obviously, the equations for $F$ and $R$ are still satisfied. Such a change would lead to a change in $T^2_{ijk}$. If $\alpha^{ij}_{k}$ changes by $\frac{a_ia_j}{a_k}$, then $(T^2)^{ij}_{k}$ changes by $\frac{a_{\bar{i}}a_{\bar{j}}a_k}{a_{\bar{k}}a_ia_j}$. Correspondingly such a change lead to the change in $T^2_i$ as
\be
T^2_i \to T^2_i\frac{a_{\bar{i}}}{a_{i}}
\ee
If $i=\bar{i}$, then $T^2_i$ does not change. If $i\neq \bar{i}$, $T^2_i$ does change. However, $T^2_i$ is not well defined for $i\neq \bar{i}$ and the degree of freedom given by $\frac{a_{\bar{i}}}{a_{i}}$ reflects exactly this arbitrariness. Note that with arbitrary $a_i$, we have
\be
(T^2_i)^*=T^2_{\bar{i}}
\ee



\end{document}